\documentclass{ article}

\usepackage{amsmath}
\usepackage{graphicx}
\usepackage{natbib}
\usepackage{subcaption}
\usepackage{comment}
\usepackage{enumitem}

\usepackage{PRIMEarxiv}

\usepackage[utf8]{inputenc} % allow utf-8 input
\usepackage[T1]{fontenc}    % use 8-bit T1 fonts
\usepackage{hyperref}       % hyperlinks
\usepackage{url}            % simple URL typesetting
\usepackage{booktabs}       % professional-quality tables
\usepackage{amsfonts}       % blackboard math symbols
\usepackage{nicefrac}       % compact symbols for 1/2, etc.
\usepackage{microtype}      % microtypography
\usepackage{lipsum}
\usepackage{fancyhdr}       % header
\usepackage{graphicx}       % graphics
\graphicspath{{media/}}     % organize your images and other figures under media/ folder
\newtheorem{thm}{Theorem}

\def\bt{\begin{thm}}
\def\et{\end{thm}}

\def\be{\begin{eqnarray}}
\def\ee{\end{eqnarray}}
\def\bes{\begin{eqnarray*}}
\def\ees{\end{eqnarray*}}

\def\bfbeta{{\boldsymbol\beta}}
\def\cF{{\cal F}}
\def\parrow{\buildrel{ p}\over\longrightarrow}

\def\bfZ{{\bf Z}}
\def\bfz{{\bf z}}
\def\bfY{{\bf Y}}
\def\bfy{{\bf y}}
\def\Var{{\rm Var}}

\def\AVar{{\rm AVar}}
\def\ACov{{\rm ACov}}
\def\cI{{\cal I}}

\def\bd{\begin{description}}
\def\ed{\end{description}}

\newcommand{\Eq}[2]{\begin{#1}#2\end{#1}}

\title{Covariate-adjusted Group Sequential Comparisons of Survival Probabilities
}

\author{
  Peter Zhang, Brent Logan, Michael Martens \\
  Division of Biostatistics \\
  Medical College of Wisconsin \\
  Milwaukee, Wisconsin, U.S.A
}

\begin{document}
\maketitle

\begin{abstract}
In confirmatory clinical trials, survival outcomes are frequently studied and interim analyses for efficacy and/or futility are often desirable. 
Methods such as the log rank test and Cox regression model are commonly used to compare treatments in this setting. They rely on a proportional hazards (PH) assumption and are subject to type I error rate inflation and loss of power when PH are violated.  Such violations may be expected a priori, particularly when the mechanisms of treatments differ such as immunotherapy vs. chemotherapy for treating cancer.
We develop group sequential tests for comparing survival curves with covariate adjustment
that allow for interim analyses in the presence of non-PH and offer easily interpreted, clinically meaningful summary measures of the treatment effect.
The joint distribution of repeatedly computed test statistics
converges to the canonical joint distribution with a Markov structure.
The asymptotic distribution of the test statistics allows marginal comparisons of survival probabilities
at multiple fixed time points and facilitates both critical value specification to maintain type I error control and sample size/power determination.
Simulations demonstrate that the achieved type I error rate and power of the proposed tests meet targeted levels and are robust to
the PH assumption and covariate influence.
The proposed tests are illustrated using a clinical trial dataset from the Blood and Marrow Transplant Clinical Trials Network 1101 trial.
\end{abstract}

% keywords can be removed

\section{Introduction}

Group sequential ({\bf GS}) designs are a popular class of adaptive designs in which one or more interim analyses
are performed at preplanned milestones during a clinical trial \citep{JennTurn99}.
These multiple interim evaluations of accumulating data
offer potential benefits including a reduction in the resources and monetary costs of the trial
and the opportunity for new and efficacious treatments to be more rapidly identified and made available to patients.
The GS method has the greatest impact on the late stage of a clinical trial program
(i.e., phase III studies), where the sample sizes of trials are typically large
and the benefits of early termination of the trial are most substantial \citep{Todd2007}.
Although the GS design can clearly be advantageous over a fixed sample design,
where no interim analyses are conducted,
GS trials need to be performed with care to avoid inflation of the type I error rate
due to taking multiple looks at trial data.
To preserve the overall type I error probability at the prespecified level,
stopping rules for interim and final analyses need
to be obtained based on the joint distribution of the test statistics from the different analyses.
For GS analysis methods whose test statistics follow the canonical joint distribution,
%discussed in ,
the critical values to use at each analysis
%are straightforward to calculate
can be calculated with standard methods \citep{JennTurn99},
%Standard methods are available for designing GS trials with analysis procedures that follow this distribution,
including Pocock and O’Brien-Fleming designs \citep{Poco77, OBriFlem79}
and error spending functions \citep{LanDeMe83}.

Clinical trials often involve the statistical analysis of time-to-event data
%such as progression-free survival (PFS) or overall %survival (OS) 
to determine the benefit of a treatment or therapy, particularly in oncology research.
The log-rank test \citep{Mant66}, the weighted log-rank test \citep{Yang19},
and the Cox proportional hazards ({\bf PH}) model \citep{Cox72} are widely used %statistical tools 
for data analysis 
in GS clinical trials with survival endpoints.
The log-rank test is commonly adopted for comparing treatment groups without covariate adjustment
and 
is well known to be efficient %against proportional hazards alternatives, 
%i.e., 
when survival curves have PH at all time points \citep{Flem11}.
In the presence of baseline covariates, the hazard ratio (HR) under the Cox PH model
is widely used to quantify treatment effects. 
When the ratio of the hazard functions of the two treatment groups is constant over time,
the HR can be conveniently interpreted as the relative instantaneous risk.
%quantifying the reduction in hazard induced by the interventional treatment versus control.
However, when the HR is nonproportional over time, 
a Cox PH model may lead to incorrect conclusions 
because the resulting estimates of the HR depend on the follow-up duration and censoring distribution;
the interpretation of the HR estimate becomes less useful clinically in this context as well 
\citep{Saad18, Tian18, Zhao19}.
%(Saad et al., 2018; Tian et al., 2018; Zhao et al., 2019).
Nonproportional hazards ({\bf NPH}) can occur when the survival curves of the two treatment 
arms cross or when one therapy has a delayed or diminishing treatment effect compared to the other.
The crossing survival curves situation can occur when treatments operating through differing modalities are assessed;
for instance, comparisons of surgical versus nonsurgical treatments have observed this pattern \citep{Howa97}. Moreover, clinical trials studying immune checkpoint inhibitors (ICIs) and other cancer immunotherapies have observed NPH when contrasting these treatments to chemotherapy and radiation therapy, including delayed treatment effects \citep{KantHiga10, VenoNied17} and crossing survival curves \citep{BorgPaz15}. \cite{Chen13}, \cite{FerrPilo18}, and \cite{Hoos12} discussed the scientific rationale behind the existence of these patterns in immunotherapy studies and recommended that analysis methods other than the log-rank test and Cox model be used in this context. Another case is illustrated by the Blood and Marrow Transplant Clinical Trials Network (BMT CTN) 1101 trial, which compared progression free survival (PFS) between two types of transplant. During the design stage of this trial, there were concerns about the possibility of NPH between treatment groups due to the different risk profiles of the transplant methods under study, and so a Kaplan-Meier based analysis was employed to compare PFS at 2 years post-randomization.

As an alternative summary measure to the HR for describing the magnitude of the treatment effect,
the difference in survival probabilities ({\bf SPs}) at a fixed follow-up time has been proposed 
to quantify the difference in survival outcomes between treatment arms.
The SP measure remains clinically interpretable even when survival curves have NPH,
facilitating communications between physicians and patients.
GS tests have been developed for comparing SPs; see, for example, \cite{LinShen96, Loga15}.
However, these GS tests do not allow adjustment for baseline covariates that may impact survival time. \cite{GailByar86} and \cite{ZhanLobe07} proposed methods for performing covariate-adjusted comparisons of SPs using stratified Kaplan-Meier estimates and a treatment-stratified Cox model; these techniques have not been studied in the group sequential setting, though.

Randomization yields balanced and comparable treatment groups on average with respect to covariate distributions.
However, since a single clinical trial is subject to stochastic phenomena,
it may exhibit some level of nontrivial covariate imbalance
for which an analysis without covariate adjustment
may compromise the aim of accomplishing an unbiased and statistically efficient treatment comparison.
\cite{CiolMart13} and \cite{CiolMart14} have shown that,
when a baseline covariate impacts the outcome of interest,
an unadjusted test for the treatment effect in a randomized GS trial
can suffer from an inflated type I error rate, diminished power and conditional power,
and bias in treatment effect estimation when covariate values differ between treatment groups;
these issues can still arise in large trials.
This emphasizes the need for covariate adjustment in randomized clinical trials.
According to \cite{PocoAssm02},
when the baseline covariates are highly predictive of the survival outcomes,
covariate adjustment can also increase the precision of the estimated treatment difference,
thereby improving the power to detect a treatment effect in a clinical trial.

The current biostatistical literature lacks an analysis method
that incorporates both GS testing and covariate adjustment for direct comparisons of SPs.
This paper proposes an innovative GS test to account for NPH settings using covariate-adjusted estimates of SPs
under a treatment-stratified Cox PH regression model, extending the method of \cite{ZhanLobe07} to the GS setting.
This offers a robust yet effective alternative to a Cox model-based comparison of treatments for situations where proportionality is suspected or expected to be violated.  It was previously shown that, if an analysis method is semiparametric efficient for the target parameter, its test statistics across analyses will asymptotically follow the canonical joint distribution \citep{SchaTsia97}. The stratified Cox model is known to be semiparametric efficient for estimating its log hazard ratio parameters \citep{ZengLin06}. However, because the parameter of interest in the method of \cite{ZhanLobe07} is a difference in SPs, each of which is a function of both the log hazard parameters and the infinite dimensional, unspecified baseline hazard function of the Cox model, these findings do not imply that the test statistics from differences in adjusted SPs will also follow the canonical distribution in a GS setting. Therefore, we employ a direct approach via martingale theory to derive the large sample distribution of these test statistics, showing that they follow the canonical joint distribution asymptotically. This allows marginal comparisons of adjusted SPs between treatment arms 
at multiple fixed time points for many different trial settings, using readily available methods to determine critical values to meet type I error rate and power specifications.

This paper is organized as follows. 
Section 2 proposes a GS test for treatment comparisons via covariate-adjusted SPs and
studies the asymptotic behaviors of repeatedly computed test statistics over time.
Section 3 presents a simulation study that investigates 
the finite-sample performance of the proposed GS test in terms of meeting type I error rate
and power specifications for a randomized group sequential trial. 
Section 4 applies the covariate-adjusted test to a real clinical trial dataset from the BMT CTN 1101 study.
We conclude with a discussion in Section 5.

\section{Methods}
\label{sec:methods}

\subsection{Survival Data Notation}

Consider a randomized clinical trial designed to compare a survival endpoint between two treatment groups, a 
control group indexed by $i = 0$ and an intervention group indexed by $i = 1$.
The $i^{th}$ treatment group is composed of $n_i$ individuals with $n=n_0+n_1$ being the total sample size.
In many experiments, subjects under study may be recruited over a period of time.
Suppose that staggered enrollment into the trial is occurring under a group sequential study
design, such that patients are gradually entering the study at differing calendar times.
Because patient data are accumulating over calendar time, 
we need to consider what information will be available for each patient at a given calendar time of interim analysis.
Thus, there are two time scales %to be accounted for
 in group
sequential tests; one is the survival time, denoted by $t$, and the other one is the calendar time, denoted by $u$. Assume that the calendar age of the study is bounded above by $\tau$.

For the $j$th individual in group $i$, let $E_{ij}$ denote the calendar time of enrollment when the
individual is randomized and let $T_{ij}$ and $C_{ij}$  
be the time from entry to event and
time from entry to right censoring, respectively. 
Thus, the $n_i$ individuals in group $i$ enter the trial at times $E_{i1}, \ldots, E_{in_i}$
and have event times $T_{i1}, \ldots, T_{in_i}$, possibly unobserved, measured from time of entry. 
If the data are analyzed at calendar time $u$, 
then individual $j$ in group $i$ will be censored if $T_{ij} > u - E_{ij}$.
Effectively, the calendar time $u$ serves as a form of administrative censoring for patients who enrolled before $u$.
However, censoring may also occur randomly from other causes, 
and $C_{ij}$ represents the corresponding potential random right-censoring time for individual $j$ in group $i$.
Denote $a^+= \max\{0, a\}$. % and $a^-= \max\{0, -a\}$.
 
The observed data in group sequential testing problems with right-censored failure time data are
independent observations $\{X_{ij}(u), \delta_{ij}(u), \bfZ_{ij}, i = 0,1, j = 1, \ldots, n_i\}$,
where $X_{ij}(u) = \min\{T_{ij}, C_{ij}, (u-E_{ij})^+\}$ represents the follow-up time 
observed for individual $j$ in group $i$ at calendar time $u$
and is the minimum of the failure time, random censoring time, and calendar time from entry;
$\delta_{ij} = I(T_{ij} \leq \min \{C_{ij}, (u-E_{ij})^+\})$ 
is the indicator that an event has been observed by calendar time $u$;
and $\bfZ_{ij} = (Z_{ij1}, \ldots, Z_{ijp})^T$ is a $p \times 1$ column vector of baseline covariates. 

For survival time $t \geq 0$,
let $S_i(t)=P(T_{ij} > t)$ denote the survival function
under treatment arm $i$ with $i=0,1$.
Given a fixed time point $t_0$,
$S_i(t_0)$ represents the probability that an individual in treatment group $i$ will survive beyond time $t_0$.
The distribution of $T_{ij}$ may depend on a vector of pre-treatment covariates $\bfZ_{ij}$.
Without loss of generality, we assume that the final analysis occurs at calendar time $u_K$, where $u_K \le \tau$.

\subsection{Covariate-adjusted Survival Probabilities with Staggered Entry}
\label{s:ASP}

We aim to compare survival probabilities between treatments 0 and 1 using group sequential testing
with covariate adjustment using the method of \cite{ZhanLobe07}, which employed a treatment-stratified Cox PH model for adjusting survival probabilities in the fixed sample design setting.
The use of adjustment for baseline covariates
is often fruitful both in making allowance for chance covariate imbalance between treatment arms and in improving power to detect a treatment effect
by exploiting information provided through the covariate and hazard rate relationships.
We focus on group sequentially testing the SP difference, the parameter quantifying the treatment effect.

Specifically,
we adjust the comparison of SPs
between two treatment arms for the effects of baseline covariates
through use of the following treatment-stratified Cox PH regression model
\citep{Kalb11, Zuck98}
%(Kalbfleisch and Prentice, 2002; Zucker, 1998)
with group status serving as the stratification variable:
\be
\lambda_i(t|\bfZ) = \lambda_{0i}(t) \exp({\bfbeta}^T \bfZ) ~~
\hbox{for treatment arm}~ i=0,1 ~~\hbox{and}~~ 0 \leq t \leq \tau,
\ee
where $\lambda_{0i}(t)$ represents an unknown treatment-specific baseline hazard function of $T_{ij}$ 
for treatment group $i$
and $\bfbeta = (\beta_1, \ldots, \beta_p)^T$ is a column vector of parameters to be estimated.
For group $i=0,1$, the treatment-specific baseline cumulative hazard function is denoted as
$\Lambda_{0i}(t) = \int_0^t \lambda_{0i}(s)ds$
and the corresponding treatment-specific baseline survival function is given by
$S_{0i}(t) = \exp\{-\Lambda_{0i}(t)\}$.
We assume throughout that $\lambda_{0i}(t)$ is continuous for $i=0,1$. This model assumes proportional effects on the hazard for covariates, but avoids this assumption on the treatment effect through stratification by group. 

%Under model (1),
%the log partial likelihood of $\bfbeta$ at calendar %time $u$ is given by %the product of 
%\bes
%\ell(\bfbeta,u) %= \log L(\bfbeta,u) 
%= \sum_{i=0}^1\sum_{j=1}^{n_i} \delta_{ij}(u) %\biggl[\bfbeta^T \bfZ_{ij}
%- \log \biggl\{\sum_{k \in R_{ij}(u)} \exp (\bfbeta^T %\bfZ_{ik})\biggr\}\biggr].
%\ees
%the treatment-specific partial likelihoods, one %arising from each treatment group,
%\bes
%L(\bfbeta,u) 
%= \prod_{i=0}^1\prod_{j=1}^{n_i} \biggl\{{\exp %(\bfbeta^T \bfZ_{ij}) \over
%\sum_{k \in R_{ij}(u)} \exp (\bfbeta^T \bfZ_{ik})}%\biggr\}^{\delta_{ij}(u)},
%\ees
%where $R_{ij}(u) = \{k:~ 1 \leq k \leq n_i, ~ X_{ik}(u) \geq X_{ij}(u)\}$
%is the set of study subjects at risk of failure just prior to $X_{ij}(u)$ for $i=0,1$ and $j = 1, \ldots, n_i$.
%The corresponding log partial likelihood of $\bfbeta$ %is given by

Under model (1),
the maximum partial likelihood estimator (MPLE) $\hat\bfbeta(u)$ of $\bfbeta$ is obtained by finding the root of 
 the partial score function at calendar time $u$, % which is written as
\iffalse
\begin{equation}
{
U(\bfbeta,u)
%&= {\partial \ell(\bfbeta,u) \over \partial \bfbeta}
%= \sum_{i=0}^1 \sum_{j=1}^{n_i} \delta_{ij}(u)
%\biggl\{\bfZ_{ij} - {\sum_{k \in R_{ij}(u)} \bfZ_{ik} %\exp (\bfbeta^T \bfZ_{ik}) \over
%\sum_{k \in R_{ij}(u)} \exp (\bfbeta^T \bfZ_{ik})}%\biggr\} \cr
=\sum_{i=0}^1 \sum_{j=1}^{n_i} \int_0^u \bigl\{\bfZ_{ij} - E_i(\bfbeta,u,s)\bigr\} N_{ij}(u,ds), 
}
\end{equation}
\fi
$U(\bfbeta,u)
=\sum_{i=0}^1 \sum_{j=1}^{n_i} \int_0^u \bigl\{\bfZ_{ij} - E_i(\bfbeta,u,s)\bigr\} N_{ij}(u,ds)$,
where $N_{ij} (u,t) = I\bigl(X_{ij}(u) \leq t, \delta_{ij}(u) = 1\bigr)$
is the simple counting process for the number of failures observed in $(0, t]$
for the $j$th individual in the $i$th group at calendar time $u$,
$Y_{ij} (u,t) = I\bigl(X_{ij}(u) \geq t\bigr)$ is the corresponding at-risk process, 
\iffalse
and 
%\begin{equation}
\begin{align*}
S_{0i}(\bfbeta,u,t) &= {1 \over n_i} \sum_{j=1}^{n_i} Y_{ij}(u,t) \exp(\bfbeta^T \bfZ_{ij}), \qquad
E_i(\bfbeta,u,t) = {S_{1i}(\bfbeta,u,t) \over S_{0i}(\bfbeta,u,t)}, \cr
S_{1i}(\bfbeta,u,t) &= {1 \over n_i} \sum_{j=1}^{n_i} Y_{ij}(u,t) \exp(\bfbeta^T \bfZ_{ij})\bfZ_{ij}.
\end{align*}
%\end{equation}
%The MPLE $\hat\bfbeta(u)$
%can be obtained as a solution to the system of partial %score equations: $U(\bfbeta,u) = 0$.
\fi
$S_{0i}(\bfbeta,u,t) = n_i^{-1} \sum_{j=1}^{n_i} Y_{ij}(u,t) \exp(\bfbeta^T \bfZ_{ij})$,
$E_i(\bfbeta,u,t) = {S_{1i}(\bfbeta,u,t) / S_{0i}(\bfbeta,u,t)}$, 
and $S_{1i}(\bfbeta,u,t) = n_i^{-1} \sum_{j=1}^{n_i} Y_{ij}(u,t) \exp(\bfbeta^T \bfZ_{ij})\bfZ_{ij}$.
The partial observed information matrix at calendar time $u$ is found as
\iffalse
\begin{equation}
{
\cI(\bfbeta,u) 
= - {\partial^2 \ell(\bfbeta,u) \over \partial\bfbeta\partial\bfbeta^T}
= \int_0^u \sum_{i=0}^1 \sum_{j=1}^{n_i} V_i(\bfbeta,u,s) N_{ij}(u,ds),
}
\end{equation}
\fi
$\cI(\bfbeta,u)
= - {\partial U(\bfbeta,u) / \partial\bfbeta^T}
= \sum_{i=0}^1 \sum_{j=1}^{n_i} \int_0^u V_i(\bfbeta,u,s) N_{ij}(u,ds)$,
where
\iffalse
%\Eq{align*}
 S_{2i}(\bfbeta,u,t)
= {1 \over n_i} \sum_{j=1}^{n_i} Y_{ij}(u,t) \exp(\bfbeta^T \bfZ_{ij})\bfZ_{ij}\bfZ_{ij}^T, \quad  V_i(\bfbeta,u,t) = {S_{2i}(\bfbeta,u,t) \over S_{0i}(\bfbeta,u,t)} - E_i(\bfbeta,u,t) E_i^T(\bfbeta,u,t).
\fi
$S_{2i}(\bfbeta,u,t)
= n_i^{-1} \sum_{j=1}^{n_i} Y_{ij}(u,t) \exp(\bfbeta^T \bfZ_{ij})\bfZ_{ij}\bfZ_{ij}^T$ and  $V_i(\bfbeta,u,t) 
={S_{2i}(\bfbeta,u,t) / S_{0i}(\bfbeta,u,t)}- E_i(\bfbeta,u,t) E_i^T(\bfbeta,u,t).$
We can show that the expected partial information matrix is equal to
$I(\bfbeta,u) = E\{{\cal I}(\bfbeta,u)\} = \Var\{U(\bfbeta,u)\}.$

The following assumptions are made about the data:
\begin{enumerate}
\item For each $i=0,1$, the set of quadruples $\{(E_{ij}, T_{ij}, C_{ij}, \bfZ_{ij}),~ j=1, \ldots, n_i\}$ are independent and identically distributed;
\item The random vectors $\{(E_{0j}, T_{0j}, C_{0j}, \bfZ_{0j}),~ j=1, \ldots, n_0\}$ and
$\{(E_{1j}, T_{1j}, C_{1j}, \bfZ_{1j}),~ j=1, \ldots, n_1\}$ are independent;
\item $E_{ij}$ is independent of $(T_{ij}, C_{ij}, \bfZ_{ij})$ for $i=0,1, j=1, \dots, n_i$;
\item For each $i=0,1$, there exists $\rho_i \in (0,1)$ such that $\lim_{n \to \infty} n_i/n = \rho_i$;
\item Model (1) holds.
\end{enumerate}
Assumption (3) is similar to that made in \cite{SellSieg83} for a sequential analysis of the Cox model and presumes that the distributions of covariates and event and censoring times do not change over calendar time. 

Under model (1),
the treatment-specific baseline cumulative hazard function $\Lambda_{0i}(t)= \int_0^t \lambda_{0i}(s)ds$ can be estimated  
at analysis $k$ using the standard method of \cite{Bres74}.
Let $\hat{\bfbeta}_{(k)}=\hat\bfbeta(u_k)$
denote the maximum partial likelihood estimator of $\bfbeta$ at calendar time $u_k$ for $k=1, \ldots, K$.
Then we can obtain an estimator of $\Lambda_{0i}(t)$ at analysis $k$
in a similar manner to the estimator of $\Lambda_{0i}(t)$ in the fixed sample design setting, given by
$\hat\Lambda_{0i}(u_k,t) 
= \int_0^t \bigl\{\sum_{j=1}^{n_i} Y_{ij}(u_k,s) \exp(\hat\bfbeta_{(k)}^T\bfZ_{ij})\bigr\}^{-1} 
\bigl\{\sum_{j=1}^{n_i} dN_{ij}(u_k,s)\bigr\}$ for $t \leq u_k$.
Because 
$S_{0i}(t) = \exp\{-\Lambda_{0i}(t)\}$ for $i=0,1$,
the treatment-specific baseline survival function $S_{0i}(t)$ is estimated by
$\hat S_{0i}(u_k,t) = \exp\{-\hat\Lambda_{0i}(u_k,t)\}$
at calendar time $u_k$ for $t \leq u_k$.

Let $\Lambda_i(t|\bfZ) = \int_0^t \lambda_i(s|\bfZ)ds$ and $S_i(t|\bfz) = P(T_{ij} \geq t|\bfZ=\bfz)$ 
denote, respectively, the cumulative hazard function and the survival function for treatment group $i=0,1$ 
conditional on covariates $\bfZ$.
Under model (1),
the treatment-specific conditional cumulative hazard function is
$\Lambda_i(t|\bfZ) = \exp(\bfbeta_0^T\bfZ)\Lambda_{0i}(t)$
and the conditional treatment-specific survival function is calculated as
$S_i(t|\bfZ) = \exp\{-\Lambda_i(t|\bfZ)\} = \{S_{0i}(t)\}^{\exp(\bfbeta_0^T\bfZ)}$,
where $\bfbeta_0$ is the true value of $\bfbeta$.
At analysis $k$, $\Lambda_i(t|\bfZ)$ and $S_i(t|\bfZ)$ 
can be estimated by
$\hat\Lambda_i(u_k,t|\bfZ) =  \exp(\hat\bfbeta_{(k)}^T\bfZ)\hat\Lambda_{0i}(u_k,t)$ and
$\hat S_i(u_k,t|\bfZ) = \{\hat S_{0i}(u_k,t)\}^{\exp(\hat\bfbeta_{(k)}^T\bfZ)}$ for $t \leq u_k$ and $i=0,1$.

Since $S_i(t_0) = E\{S_i(t_0|\bfZ)\}$,
the survival probability $S_i(t_0)$ at a fixed survival time $t_0$ is a marginal survival probability for group $i$ at $t_0$ that averages the conditional survival probabilities $S_i(t_0|\bfZ)$ over all patients in the target population. 
%As in Zucker (1998), we define this probability as
%\Eq{align*}
%{
%S_i(t_0) = {1 \over n} \sum_{g=0}^1\sum_{j=1}^{n_g} S_i(t_0|\bfZ_{gj})
%= {1 \over n} \sum_{g=0}^1\sum_{j=1}^{n_g} \exp\{-\exp(\bfbeta_0^T\bfZ_{gj}) \Lambda_{0i}(t_0)\},
%}
This is naturally estimated at analysis $k$ by the covariate-adjusted survival probability
$$\hat S_i(u_k,t_0) = {1 \over n} \sum_{g=0}^1\sum_{j=1}^{n_g} \hat S_i(u_k,t_0|\bfZ_{gj})
= {1 \over n} \sum_{g=0}^1\sum_{j=1}^{n_g} \{\hat S_{0i}(u_k,t_0)\}^{\exp(\hat\bfbeta_{(k)}^T\bfZ_{gj})},$$
which is an average of survival probability estimates under treatment $i$ over all patients in the study population.
%$\hat S_i(u_k,t_0) = n^{-1} \sum_{g=0}^1\sum_{j=1}^{n_g} \hat S_i(u_k,t_0|\bfZ_{gj})
%= n^{-1} \sum_{g=0}^1\sum_{j=1}^{n_g} \{\hat S_{0i}(u_k,t_0)\}^{\exp(\hat\bfbeta_{(k)}^T\bfZ_{gj})}$
%for treatment arm $i=0,1$.

\subsection {Group Sequential Test of Treatment Effect}

A two-sample comparison of survival probabilities at a fixed time point $t_0 \in (0,\tau]$ between the control group and the intervention group
is established in the two-sided hypothesis:
$H_0: S_0(t_0) = S_1(t_0)$ versus $H_1: S_0(t_0) \not= S_1(t_0)$.
One-sided testing of the alternative hypothesis $H_1: S_0(t_0) < S_1(t_0)$ or $H_1: S_0(t_0) > S_1(t_0)$
can be considered for testing superiority or inferiority at time $t_0$ of group 1 relative to 0.
We aim to extend the covariate-adjusted survival probabilities described in Section \ref{s:ASP}
to the group sequential design setting for testing the null hypothesis based on accumulating survival data at $K$ calendar times,
denoted by $u_1 < u_2 < \cdots < u_K$. It is assumed that $u_1 \ge t_0,$ which ensures a positive probability exists of observing some events in the survival time interval $[0,t_0]$ for each analysis.

At each calendar analysis time $u_k$,
the successive estimators $\hat S_0(u_k,t_0)$ and $\hat S_1(u_k,t_0)$
of the treatment-specific survival probabilities $S_0(t_0)$ and $S_1(t_0)$
can be used
as a basis to conduct this test through the sequence of standardized statistics
%\begin{equation}
%Z(u_k,t_0) = {\hat S_1(u_k,t_0)-\hat S_0(u_k,t_0) \over \hat \sigma(u_k,t_0)},
%\qquad t_0 \leq u_k,~ k=1, \ldots, K,
%\end{equation}
$Z(u_k,t_0) = \{\hat S_1(u_k,t_0)-\hat S_0(u_k,t_0)\} / \hat \sigma(u_k,t_0)$
for $t_0 \leq u_k$ and $k=1, \ldots, K$,
where $\hat \sigma^2(u_k,t_0)$ is a consistent estimator for the asymptotic variance $\sigma^2(u_k,t_0)$ 
of the estimated SP difference $\hat S_1(u_k,t_0)-\hat S_0(u_k,t_0)$ at analysis $k$.
The expressions for $\sigma^2(u_k,t_0)$ and $\hat \sigma^2(u_k,t_0)$ are provided in and following Theorem 1 below.
The associated information level at calendar time $u_k$ is defined as
the reciprocal of the estimated variance of $\hat S_1(u_k,t_0)-\hat S_0(u_k,t_0)$, 
i.e. $I(u_k,t_0) = \hat \sigma^{-2}(u_k,t_0)$.
The validity of the group sequential design using the test statistics $\bigl(Z(u_1,t_0), \ldots, Z(u_K,t_0)\bigr)$
relies on the correct determination of the group-sequential boundaries based on information fraction
accrued at each interim look, which requires knowledge of the joint distribution of the test statistics. To permit flexibility in the choice of analysis times, whose number and timing is often not predetermined, the following theorem establishes the asymptotic joint null distribution of the stochastic process
$\bigl\{ \sqrt{n} \{\hat S_1(u,t_0)-\hat S_0(u,t_0)\}: ~ 0<t_0 \leq u \leq \tau \bigr\}$.

\bt
Let the survival time $t_0>0$ be given.
Suppose that the regularity conditions (A)-(H) in the Appendix are satisfied for $u \in [t_0, \tau]$.
Then under the stratified PH model (1)
and under the null hypothesis $H_0: S_0(t_0)=S_1(t_0)$,
the stochastic process
$\bigl\{\sqrt{n} \{\hat S_1(u,t_0)-\hat S_0(u,t_0)\}: ~ t_0 \leq u \leq \tau \bigr\}$
converges weakly to a Gaussian process $\xi$
with continuous sample paths, mean 0, and covariance function 
\Eq{align*}
{
\omega(u,v,t_0) &= E\{\xi(u)\xi(v)\}
= \sum_{i=0}^1 {1 \over \rho_i} c_{i1}^2(t_0) \gamma_i(\bfbeta_0,u \vee v,t_0) 
+ D^T(\bfbeta_0,t_0)\Sigma^{-1}(\bfbeta_0,u \vee v) D(\bfbeta_0,t_0) \\
&= E\{\xi^2(u \vee v)\} = \omega(u \vee v, u \vee v,t_0), 
}
where
\iffalse
\Eq{align}
{
%& C_i(\bfbeta,u,t)
%= {\sqrt{n} \over n_i} \sum_{j=1}^{n_i}\int_0^t {1 %\over S_{0i}(\bfbeta,u,s)}
%I\biggl(\sum_{j=1}^{n_i}Y_{ij}(u,s) > 0\biggr) M_{ij}%(u,ds), \cr
& \gamma_i(\bfbeta,u,t) = \int_0^t {1 \over s_{0i}(\bfbeta,u,s)} d\Lambda_{0i}(s), \qquad
Q_i(\bfbeta,t) = \int_0^t e_i(\bfbeta,s) d\Lambda_{0i}(s), \cr
& \Sigma(\bfbeta,u)=\sum_{i=0}^1 \rho_i \int_0^u v_i(\bfbeta,u,s)s_{0i}(\bfbeta,u,s) \lambda_{0i}(s)ds \cr
& c_{i1}(t) = \sum_{g=0}^1 \rho_g E\{S_i(t|\bfZ_{g1})\exp(\bfbeta_0^T\bfZ_{g1})\}, \cr
& c_{i2}(t) = \sum_{g=0}^1 \rho_g E\{S_i(t|\bfZ_{g1})\exp(\bfbeta_0^T\bfZ_{g1})\bfZ_{g1}\}, \cr
& D_i(\bfbeta,t) = c_{i1}(t) Q_i(\bfbeta,t) - \Lambda_{0i}(t) c_{i2}(t), 
\qquad D(\bfbeta,t) = D_1(\bfbeta,t) -D_0(\bfbeta,t). \hskip0.4in
}
\fi
$\gamma_i(\bfbeta,u,t) = \int_0^t {1 / s_{0i}(\bfbeta,u,s)} d\Lambda_{0i}(s), \quad$
$Q_i(\bfbeta,t) = \int_0^t e_i(\bfbeta,s) d\Lambda_{0i}(s),$ \\
$\Sigma(\bfbeta,u)=\sum_{i=0}^1 \rho_i \int_0^\tau v_i(\bfbeta,u,s)s_{0i}(\bfbeta,u,s) \lambda_{0i}(s)ds$,
$c_{i1}(t) = \sum_{g=0}^1 \rho_g E\{S_i(t|\bfZ_{g1})\exp(\bfbeta_0^T\bfZ_{g1})\},$
\\ $c_{i2}(t) = \sum_{g=0}^1 \rho_g E\{S_i(t|\bfZ_{g1})\exp(\bfbeta_0^T\bfZ_{g1})\bfZ_{g1}\},$
$D_i(\bfbeta,t) = c_{i1}(t) Q_i(\bfbeta,t) - \Lambda_{0i}(t) c_{i2}(t),$
and \\ $D(\bfbeta,t) = D_1(\bfbeta,t) -D_0(\bfbeta,t).$
In particular, $E\{\xi^2(u)\}=\omega(u,u,t_0) = \sigma^2(u,t_0)$.
\et
The derivation of Theorem 1 is included in the Supplementary Materials. This result constitutes a basis for the design and analysis of group sequential clinical trials
using the difference of covariate-adjusted survival probabilities.
Under the null hypothesis,
the joint distribution of 
$\bigl(\sqrt{n}\{\hat S_1(u_1,t_0)-\hat S_0(u_1,t_0)\}, \ldots, \sqrt{n} \{\hat S_1(u_K,t_0)-\hat S_0(u_K,t_0)\}\bigr)$
possesses asymptotically a normal independent (reverse) increments covariance structure in that
%\begin{align}
$
 \ACov\bigl(\sqrt{n} \{\hat S_1(u_k,t_0) - \hat S_0(u_k,t_0)\},
\sqrt{n} \{\hat S_1(u_l,t_0) - \hat S_0(u_l,t_0)\} \bigr) 
 = \AVar\bigl(\sqrt{n} \{\hat S_1(u_{k \vee l},t_0) - \hat S_0(u_{k \vee l},t_0)\}\bigr),$
%\end{align}
where ACov and AVar stand for asymptotic covariance and asymptotic variance, respectively.
This implies that, like many other group sequential analysis methods discussed in Jennison and Turnbull (2000),
the repeatedly computed Wald test statistics $Z(u_1,t_0), \ldots, Z(u_K,t_0)$
follow asymptotically the canonical joint distribution with a Markov structure, i.e.
%$$\ACov\bigl(Z(u_k,t_0),~Z(u_l,t_0)\bigr) = {\sigma(u_{k \vee l},t_0) \over \sigma(u_{k \wedge l},t_0)}.$$
\\ $\ACov\bigl(Z(u_k,t_0),~Z(u_l,t_0)\bigr) = {\sigma(u_{k \vee l},t_0) / \sigma(u_{k \wedge l},t_0)}.$
Because this asymptotic distribution has exactly the same structure 
as many commonly-used sequentially computed test statistics,
standard methodology can be used to design and analyze group sequential tests 
using the test statistics $Z(u_1,t_0), \ldots, Z(u_K,t_0)$. 
As a result, we will use the large-sample joint null distribution of $Z(u_1,t_0), \ldots, Z(u_K,t_0)$
to obtain appropriate critical values for conducting covariate-adjusted group sequential testing of SPs.
This allows the use of standard group sequential methods for early stopping.
The asymptotic variance $\sigma^2(u_k,t_0)$
can be estimated consistently by \\
\iffalse
\Eq{align*}
{
\hat\sigma^2(u_k,t_0)
&= {n \over n_0} \hat c_{01}^2(u_k,t_0) \hat\gamma_0(\hat\bfbeta_{(k)},u_k,t_0)
+ {n \over n_1} \hat c_{11}^2(u_k,t_0) \hat\gamma_1(\hat\bfbeta_{(k)},u_k,t_0) \cr
&\quad + \hat D^T(\hat\bfbeta_{(k)},t_0) \hat \Sigma(\hat\bfbeta_{(k)},u_k) \hat D(\hat\bfbeta_{(k)},t_0)%, \cr
%& \hat\omega(u_k,u_l,t_0)
%= {n \over n_0} \hat c_{01}^2(u_{k \vee l},t_0) \hat\gamma_0(\hat\bfbeta_{(k \vee l)},u_{k \vee l},t_0)
%+ {n \over n_1} \hat c_{11}^2(u_{k \vee l},t_0)
\hat\gamma_1(\hat\bfbeta_{(k \vee l)},u_{k \vee l},t_0) 
%&\quad + \hat D^T(\hat\bfbeta_{(k)},t_0) \hat\Sigma^{-1}(\hat\bfbeta_{(k \vee l)},u_{k \vee l})
%\hat D(\hat\bfbeta_{(l)},t_0), \cr
}
\fi
$\hat\sigma^2(u_k,t_0)
= \sum_{i=0}^1{n / n_i} \cdot \hat c_{i1}^2(u_k,t_0) \hat\gamma_i(\hat\bfbeta_{(k)},u_k,t_0)
+ \hat D^T(\hat\bfbeta_{(k)},t_0) \hat \Sigma(\hat\bfbeta_{(k)},u_k) \hat D(\hat\bfbeta_{(k)},t_0),$
where
%for $i=0,1$,
\iffalse
\Eq{align*}
{
 \hat c_{i1}(u_k,t_0) = {1 \over n} \sum_{g=0}^1\sum_{j=1}^{n_g}
\hat S_i(u_k,t_0|\bfZ_{gj})\exp(\hat\bfbeta_{(k)}^T\bfZ_{gj}), \cr & \hat c_{i2}(u_k,t_0) = {1 \over n} \sum_{g=0}^1\sum_{j=1}^{n_g}
\hat S_i(u_k,t_0|\bfZ_{gj})\exp(\hat\bfbeta_{(k)}^T\bfZ_{gj})\bfZ_{gj}, \cr
& \hat\gamma_i(\hat\bfbeta_{(k)},u_k,t_0)
= {1 \over n_i} \int_0^{t_0} {1 \over S_{0i}^2(\hat\bfbeta_{(k)},u_k,s)}dN_i(u_k,s), \cr & \hat Q_i(\hat\bfbeta_{(k)},u_k,t_0)
={1 \over n_i} \int_0^{t_0}{S_{1i}(\hat\bfbeta_{(k)},u_k,s) \over S_{0i}^2(\hat\bfbeta_{(k)},u_k,s)}dN_i(u_k,s), \\
&\hat\Sigma(\hat\bfbeta_{(k)},u_k)
= {1 \over n} {\cal I}\bigl(\hat\bfbeta_{(k)},u_k\bigr), \quad \hat D_i(\hat\bfbeta_{(k)},t_0) = \hat c_{i1}(u_k,t_0) \hat Q_i(\hat\bfbeta_{(k)},u_k, t_0)
- \hat\Lambda_{0i}(u_k,t_0) \hat c_{i2}(u_k,t_0), \cr
& \hat D(\hat\bfbeta_{(k)},t_0) = \hat D_1(\hat\bfbeta_{(k)},t_0) - \hat D_0(\hat\bfbeta_{(k)},t_0).
}
\fi
\\ $\hat\gamma_i(\hat\bfbeta_{(k)},u_k,t_0)
= \int_0^{t_0} {n_i^{-1} / S_{0i}^2(\hat\bfbeta_{(k)},u_k,s)}dN_i(u_k,s), \quad \hat\Sigma(\hat\bfbeta_{(k)},u_k) = n^{-1} {\cal I}\bigl(\hat\bfbeta_{(k)},u_k\bigr)$,\\
$\hat c_{i1}(u_k,t_0)$ $= n^{-1} \sum_{g=0}^1\sum_{j=1}^{n_g}
\hat S_i(u_k,t_0|\bfZ_{gj})\exp(\hat\bfbeta_{(k)}^T\bfZ_{gj}),$
\\ $\hat Q_i(\hat\bfbeta_{(k)},u_k,t_0)
=n_i^{-1} \int_0^{t_0}{S_{1i}(\hat\bfbeta_{(k)},u_k,s) / S_{0i}^2(\hat\bfbeta_{(k)},u_k,s)}dN_i(u_k,s),$ \\
$\hat c_{i2}(u_k,t_0) = n^{-1} \sum_{g=0}^1\sum_{j=1}^{n_g}
\hat S_i(u_k,t_0|\bfZ_{gj})\exp(\hat\bfbeta_{(k)}^T\bfZ_{gj})\bfZ_{gj},$
%$\hat\gamma_i(\hat\bfbeta_{(k)},u_k,t_0)
%= \int_0^{t_0} {n_i^{-1} \over S_{0i}^2(\hat\bfbeta_{(k)},u_k,s)}dN_i(u_k,s),$
%$\hat Q_i(\hat\bfbeta_{(k)},u_k,t_0)
%=n_i^{-1} \int_0^{t_0}{S_{1i}(\hat\bfbeta_{(k)},u_k,s) \over S_{0i}^2(\hat\bfbeta_{(k)},u_k,s)}dN_i(u_k,s),$
\\
$\hat D_i(\hat\bfbeta_{(k)},t_0) = \hat c_{i1}(u_k,t_0) \hat Q_i(\hat\bfbeta_{(k)},u_k, t_0)
- \hat\Lambda_{0i}(u_k,t_0) \hat c_{i2}(u_k,t_0),$
and \\
$\hat D(\hat\bfbeta_{(k)},t_0) = \hat D_1(\hat\bfbeta_{(k)},t_0) - \hat D_0(\hat\bfbeta_{(k)},t_0)$
for $i=0,1$.

\section{Simulations}
\label{sec:simulations}
Simulation studies were conducted with the following objectives: (1) investigate the performance of the proposed test in
 meeting type I error and power specifications with practical sample sizes, and (2) compare the achieved type I error rate and power from the proposed method to existing methods.
 The survival probability at a pre-specified time point was compared between treatment groups using the proposed method,
 Kaplan-Meier estimation,  which does not adjust for covariates, 
 and the Cox PH model,
 which assumes proportionality between treatment groups. 
 %We assume there are two groups, treatment and control; let $Z_1 = I(\mbox{group = treatment})$.
%We also assume there is one influential covariate, $Z_2$. 
We simulated randomized group sequential trials to compare the treatment effect between an experimental treatment group and a control group by testing $H_0: {S}_0(\tau) = {S}_1(\tau)$ for the proposed test and Kaplan-Meier based test. For the Cox PH model based test, we included the treatment indicator $Z_W = I(\mbox{group = treatment})$ and covariates $\mathbf{Z}$ in a Cox model and evaluated $H_0^*: \beta_W = 0$, where $\beta_W$ is the parameter for $Z_W$.

Subjects were randomized between the treatment and control groups at a 1:1 ratio
and were uniformly enrolled during an accrual period [0, A].
% in which patients were enrolled into the trial uniformly.  
Let the length of the study be $L = \tau + A$.  
Group sequential tests were performed at a significance level of $5\%$ 
using the proposed test with three interim analyses.
We used the alpha spending function $0.05\min{\{1, IF^3}\}$, 
where $IF$ is the information fraction, i.e.,
the fraction of total information for the trial.
Calendar times for the three interim analyses were pre-specified so that
the expected information fraction at each analysis would be $0.50, 0.75$, and $1$. 
The total information level and interim analysis times were computed using Monte Carlo estimation
for each simulation scenario. 

Trial data were simulated as follows. 
Assume the event time for each subject follows a Weibull distribution with survival function
$S(t|Z_W, \mathbf{Z}) = \exp(-\gamma t^{\alpha})$, 
where the shape and rate parameters $\alpha$ and $\gamma$ depend on $Z_W$ and $\mathbf{Z}$ through the forms
$\alpha = \alpha_0 + \alpha_1Z_W$ and 
 $\gamma = \gamma_0 \exp(\beta_WZ_W + \bfbeta^T \bfZ)$. 
The hazard function for the Weibull distribution has the form
 $\lambda(t | Z_W, \bfZ) =\alpha\gamma t^{\alpha - 1} = (\alpha_0 + \alpha_1Z_W) \gamma_0 \exp{(\beta_WZ_W + \bfbeta^T \bfZ)}t^{\alpha_0 + \alpha_1Z_W - 1}$.       %check
% $\lambda(t | Z_W = 0, \bfZ) = \alpha_0 \gamma_0 \exp{(\bfbeta^T \bfZ)}t^{\alpha_0  - 1}$ 
%and $\lambda(t | Z_W = 1, \bfZ) =\alpha\gamma t^{\alpha - 1}$ \break $ = (\alpha_0 + \alpha_1) \gamma_0 %\exp{(\beta_W + \bfbeta^T \bfZ)}t^{\alpha_0 + \alpha_1 - 1}$,
%and 
Then, the hazard ratio of the treatment groups has the form $\lambda(t | Z_W = 1, \bfZ) /  \lambda(t | Z_W = 0, \bfZ) = (\alpha_0 + \alpha_1) / \alpha_0 \cdot \exp{(\beta_W)}t^{\alpha_1}$. 
Note that the hazard ratio equals $\exp{(\beta_W)}$ when $\alpha_1 =0$, 
and that there are PH in this case;
when $\alpha_1 \not=0$, the PH assumption is violated.
Furthermore, under $H_0$, 
 it can be shown that $\beta_{W} = \log(\tau^{-\alpha_{1}})$.

Simulation parameters were chosen to be:
 equal sample sizes per treatment group of $n_0 = n_1 = 200$ or 400;
 $\tau = 1$ or 3 years; 
either no random censoring, or random censoring following a $\exp(-\log(0.95))$ distribution to yield 5\% censoring per year;
accrual period A = 2 or 4. 
$\beta_{W}$ was set to $\log(\tau^{-\alpha})$ under $H_0$; under the alternative hypothesis $H_1: S_0(\tau) = S_1(\tau) - \delta$, $\beta_{W}$ was set to $\beta_{\delta}$,
the value of $\beta_{W}$ which corresponds to the treatment effect $\delta = S_1(\tau) - S_0(\tau)$
 at which the proposed test should have 80\% power.
 We consider two possible distributions for the set of influential covariates, $\mathbf{Z}$:
 a single covariate following the standard normal distribution,
 and two covariates following standardized Bernoulli(0.3) and Bernoulli(0.5) distributions. The covariate effects were specified as $\beta_j = \phi / \sqrt{p}$ for each covariate distribution, where $p$ is the number of covariates and $\phi$ is a parameter controlling the level of covariate influence; under this specification, the variance of the linear predictor $\bfbeta^T \bfZ$ is the same under both covariate distributions for a given value of $\phi$. Possible choices of $\phi$ were $0$, $\log(1.5)$, and $\log(2)$.
  These simulation parameter settings yield 96 possible parameter specifications for each of the treatment group and control group. 
%Three stage group sequential design employed with alpha spending function $\alpha(t) = 0.05 \min\{1,t^3\}$
% at times where we expect to have obtained 50\%, 75\%, and 100\% of the required information
We performed 10,000 simulated trials for each parameter specification and obtained
Monte Carlo estimates to evaluate the empirical type I error rate and power of the proposed tests.  
%We assessed
%the performance of the proposed tests with regards to type I %error rate, power, and
%conditional power using Monte Carlo estimates obtained from %10,000 simulated trials
%for each specification.

%\subsection{Simulation Results}

\begin{figure}[h]
 \caption{Cumulative Type I Error and Power, NPH Setting. Dashed lines indicate nominal stagewise type I error rate and power values under asymptotic distribution.}
  \centering
%\raggedright
  \includegraphics[width=8cm]{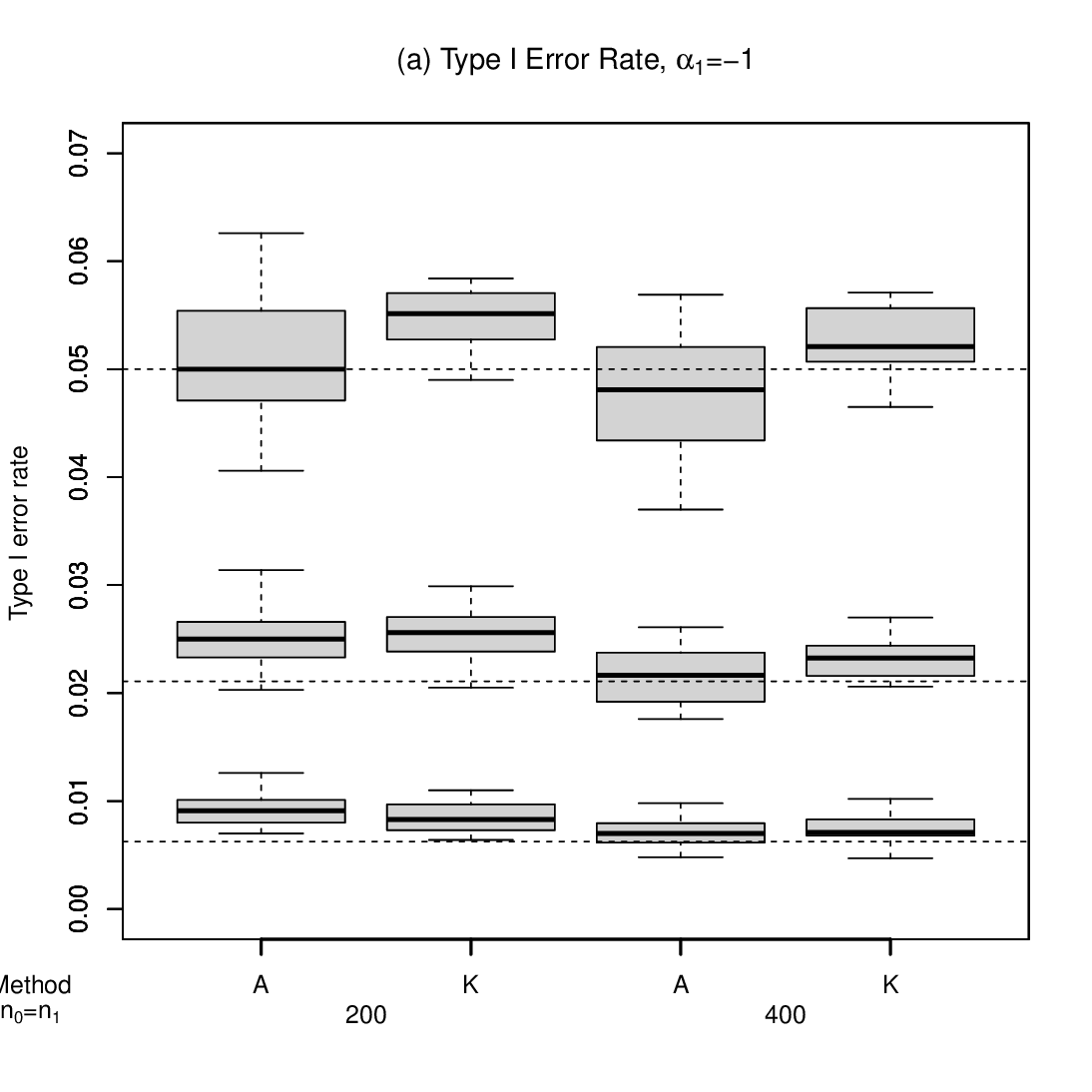}
  \includegraphics[width=8cm]{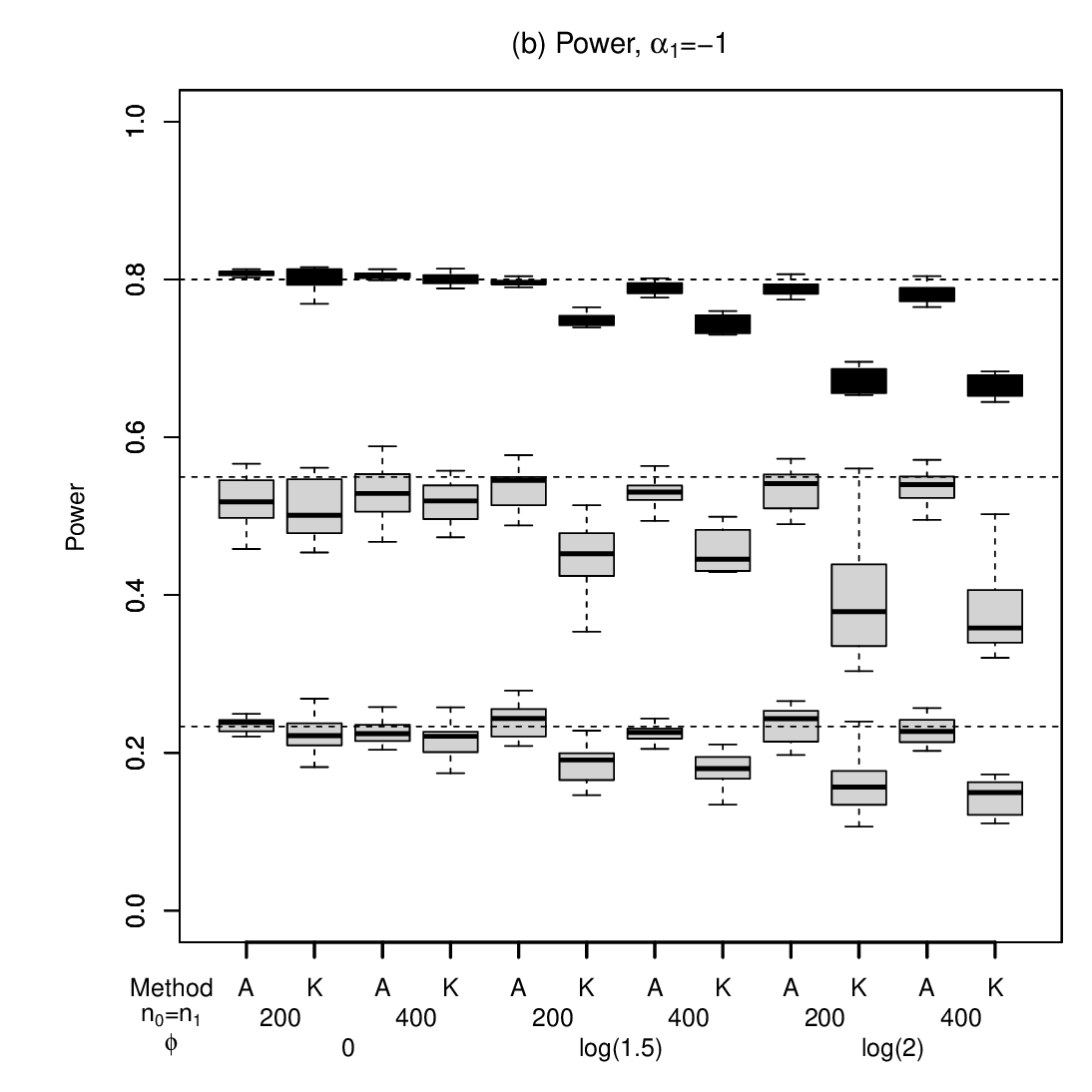}
  \vfill
   { \footnotesize A - Adjusted Survival Probability, 
  K - Kaplan-Meier}
 % \includegraphics[width=8cm]{error3.pdf}
 %  {\tiny Sample Size}%
%Figure 1b
\end{figure}

%Figure 2
%\vspace{0.4cm}
\begin{figure}[h]
 \caption{Cumulative Type I Error and Power, PH Setting. Dashed lines indicate nominal stagewise type I error rate and power values under asymptotic distribution.}
  \centering
%\raggedright
  \includegraphics[width=8cm]{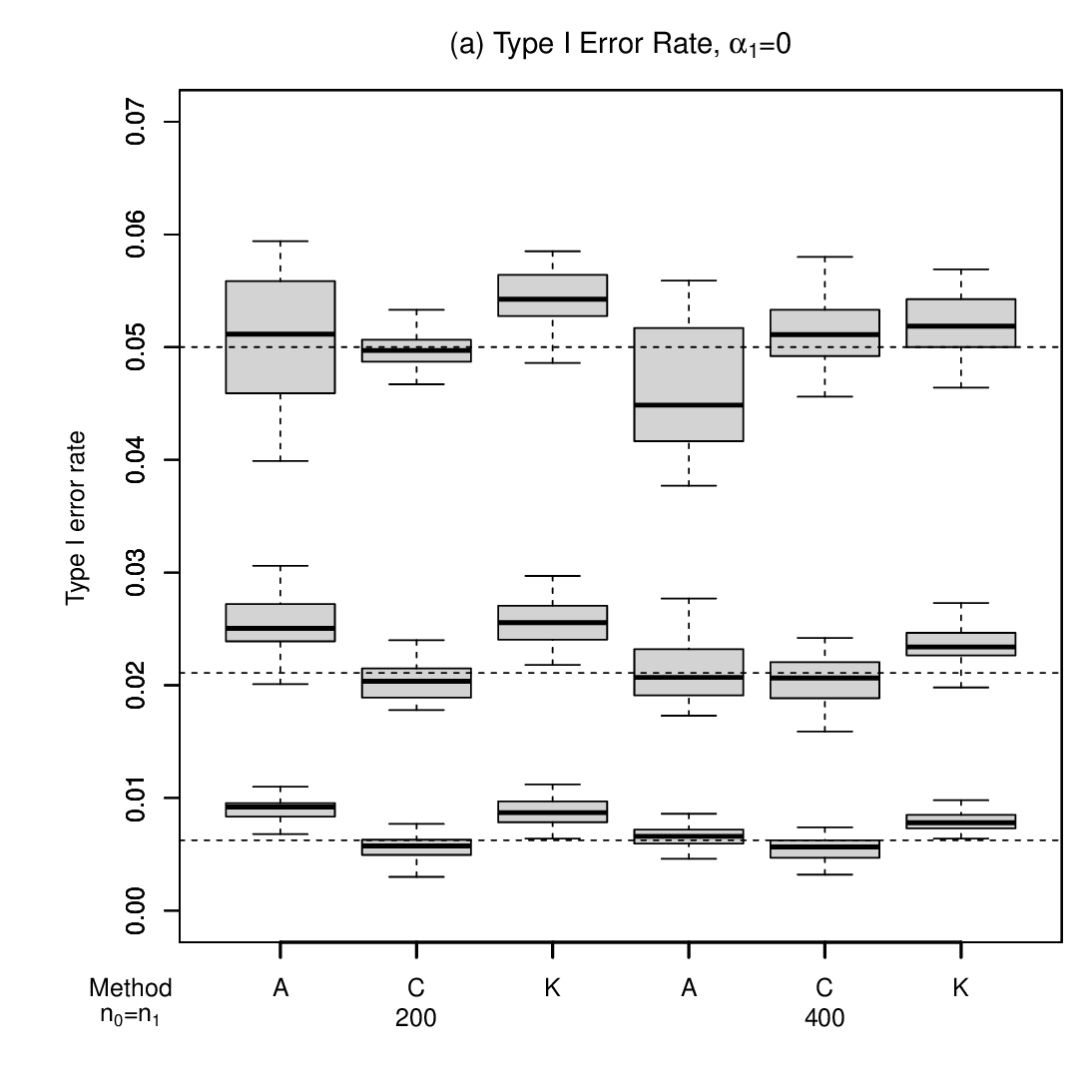}
  \includegraphics[width=8cm]{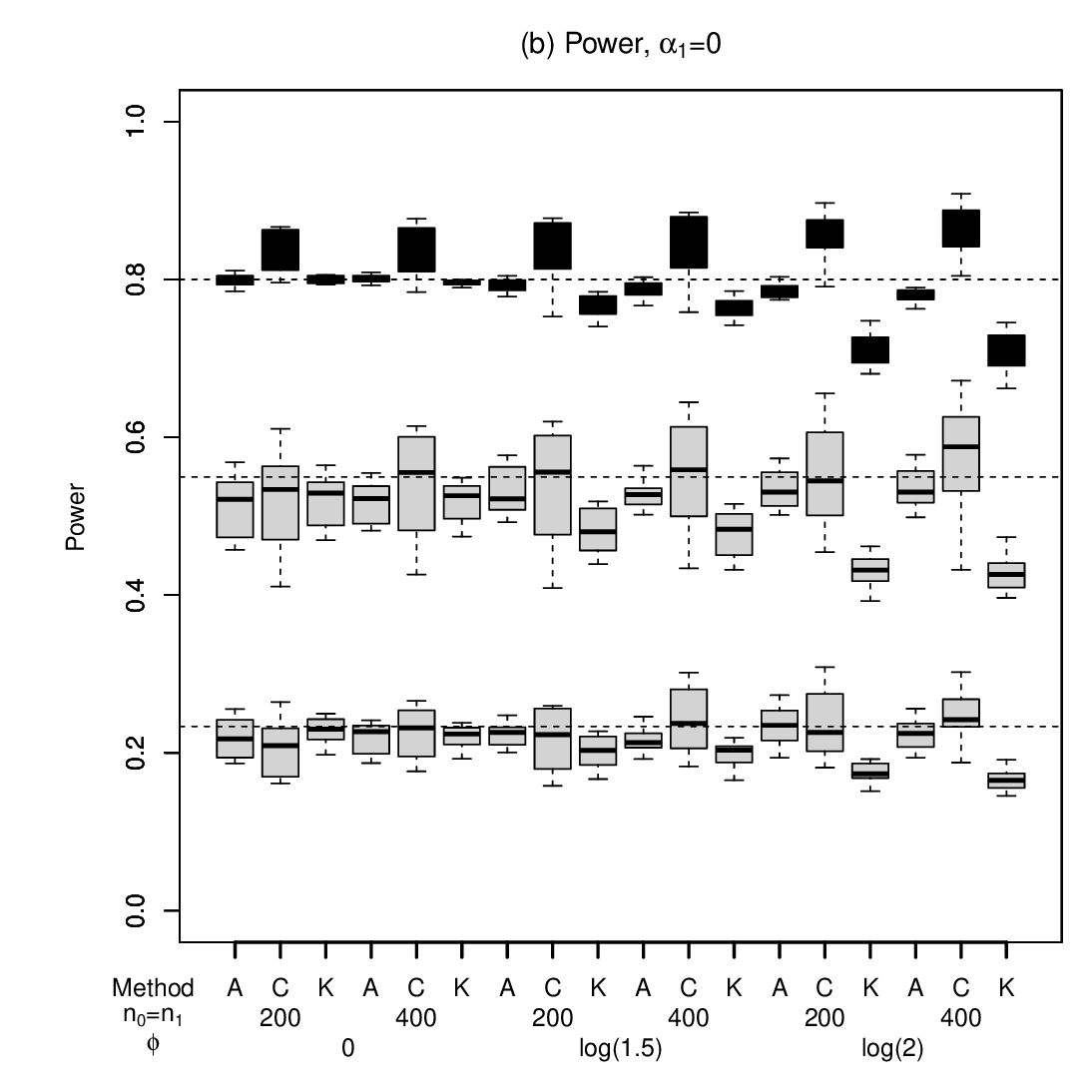}
  \vfill
  { \footnotesize A - Adjusted Survival Probability, C - Cox Model,
  K - Kaplan-Meier}
 % \includegraphics[width=8cm]{error3.pdf}
 %  {\tiny Sample Size}%
%Figure 1b
\end{figure}

Estimates of the cumulative type I error rate and power at each of the three stages 
were obtained for each method. 
Figures 1a and 1b summarize achieved stagewise type I error rates and power, respectively, for the proposed method and Kaplan-Meier based test under
the NPH setting $\alpha_1 = -1$. 
We observed the following trends.
Stagewise empirical type I error of the proposed method were lower than those of the Kaplan-Meier based analysis,
and met the targeted type I error rate of $5\%$. 
%At stage 3, the empirical type I error of the proposed %method using adjusted survival probabilities 
%meets the targeted $\alpha = 5\%$ target for both n = %200 and n = 400,
%while the empirical type I error of the Kaplan-Meier %based estimates
%are slightly above 5\% for both sample sizes.
In Figure 1b, stagewise empirical power of the proposed method were higher than those of the Kaplan-Meier based analysis for non-zero values of $\phi$;
adjusting for the covariate(s) increases power compared to
not adjusting. In addition, the proposed test is more robust to covariate effect size than the test using a Kaplan-Meier based analysis. The Cox model produced type I error rates of 70\% or higher for all scenarios in the NPH setting and, unexpectedly, had type I error rates exceeding its corresponding power levels; given its abysmal performance in this setting, 
we omitted type I error and power for the Cox model from Figure 1.

Figures 2a and 2b show achieved stagewise type I error and power, respectively, for the three methods under the PH setting ($\alpha_1 = 0$).
Similar trends were found in comparing type I error rates and power between the proposed method and Kaplan-Meier based analysis for the NPH setting. 
In the PH setting, 
the Cox model based test provided a modest boost in power
compared to the proposed method 
and met the targeted type I error rate of $5\%$.

\section{Example}
\label{s:example}
Concerns about the possibility of NPH were factored into the design of 
the Blood and Marrow Transplant Clinical Trials Network (BMT CTN) trial 1101,
a randomized multicenter phase 3 trial \citep{Fuchs2020}.
In patients with poor prognosis or who experience relapse of hematologic malignancies, 
allogeneic bone marrow or blood stem cell transplant can be a
curative treatment option.
A human leukocyte antigen (HLA)-matched unrelated donor or sibling is favored since HLA mismatches
between donor and recipient are associated with a higher chance of
graft-versus-host-disease (GVHD) and non-relapse mortality as well as inferior survival.
However, for some patients a suitable HLA-matched donor may not be available.
In these situations, alternative donor options include unrelated donor umbilical cord blood 
and an HLA-mismatched relative.
These methods motivated
the BMT CTN 1101 trial, which
compared double umbilical cord blood (UCB) and haploidentical (Haplo) related donor transplantation
for the treatment of leukemia and lymphoma in adults.
The primary endpoint was progression-free survival (PFS) and key secondary endpoints were overall survival (OS),
 non-relapse mortality, and malignancy relapse/progression.
Because investigators had concerns that the treatment groups may have NPH,
the analysis of PFS employed a fixed time point comparison
of survival probabilities at 2 years post-randomization.
Furthermore, since the trial was expected to have a 6 year duration,
a GS design was used with 3 planned interim efficacy analyses.
The results from this
trial indicate that
(1) there is no significant difference in PFS between umbilical cord blood (UCB)
and Haplo transplantation for leukemia or lymphoma
and (2) Haplo transplantation provides lower non-relapse mortality and superior OS compared to UCB transplantation.

We re-visit the 1101 trial data by re-analyzing the data using the proposed methods and comparing the results with those of the original analysis.
%and develop GS tests for comparing two treatment arms %in the presence of nonproportional hazards
%based on the difference between two survival %probabilities at a fixed yet arbitrary time point.
%We demonstrate the potential impact of the proposed GS %methods by re-analyzing the BMT CTN 1101 trial,
%which did not account for covariate adjustment in its %original GS analysis.
%Marrow Transplant Clinical Trials Network 1101 trial %(Fuchs et al., 2021), 
%which compared double umbilical cord blood (UCB) and %haploidentical (Haplo) related donor transplantation 
%for the treatment of leukemia and lymphoma in adults. 
%The objective of the BMT CTN 1101 CT was to compare post-transplant outcomes between UCB and Haplo treatment.
%%key secondary endpoints were overall survival,
% non-relapse mortality, and malignancy relapse/progression. 
 A total of 368 patients were randomized at a 1:1 ratio to the two treatment arms, UCB and Haplo.
 The original trial analysis of PFS employed a comparison
of survival at 2 years post-randomization by Kaplan-Meier estimates, 
 which do not adjust for covariates. The actual trial duration was 8 years, longer than originally planned, with an enrollment period of 6 years and all patients followed for 2 years post-randomization. To allow the possibility of declaring efficacy early and stopping the trial, a group sequential Kaplan-Meier based comparison with six planned interim analyses annually from years 3 to 8 
 was utilized to compare PFS between treatment groups. 
 The treatment groups did not differ significantly on PFS at any interim analyses, and hence the trial was not stopped early. At the final analysis, 2 year PFS estimates were 35\% for the UCB arm and 41\% for the Haplo arm, and did not differ significantly ($p$=0.41). OS at 2 years was found to be higher for the Haplo group at the final analysis compared to UCB (estimates 57\% and 46\% for Haplo and UCB, respectively; $p$=0.04).

 Several covariates are known or suspected to affect progression and mortality risk in the patient population of interest: age, gender, race, ethnicity, primary disease of lymphoma or leukemia, Karnofsky performance score, disease risk index, hematopoietic cell transplantation comorbidity index, and patient cytomegalovirus status at transplant.
 A previous secondary analysis showed that there was not significant evidence that the PH assumption was violated for these covariates.
 Therefore,
 we re-analyze the 1101 trial data to compare PFS and OS at 2 years post-randomization between the treatment arms with group sequential testing using both the proposed method to account for these influential covariates and a Kaplan-Meier analysis. The same interim analysis schedule is employed as designed for the original trial, with analyses annually from years 3 to 8.
 Alpha spending functions were used to allow for early stopping
for efficacy and were chosen from the power family in \cite{JennTurn99} with
$\rho =  3$. Information fractions for both PFS and OS were calculated using the observed information at each stage. For PFS, the total information was set at 450.550 as per trial protocol; for OS the total information was set at 386.066, the observed information at year 8 since the original study used a single analysis for OS at the end of study and did not specify a target information level for this endpoint. Note that the information fractions are different for the two methods, leading to different critical values from the alpha spending function.

\begin{figure}[h]
 \caption{Group sequential test statistics and critical values, progression-free survival. Dashed lines indicate critical values; solid lines indicate test statistics. Neither the proposed method's nor the Kaplan-Meier based comparison's test statistics exceeded the critical values at any stage.}
  \centering
%\raggedright
  \includegraphics[width=8cm]{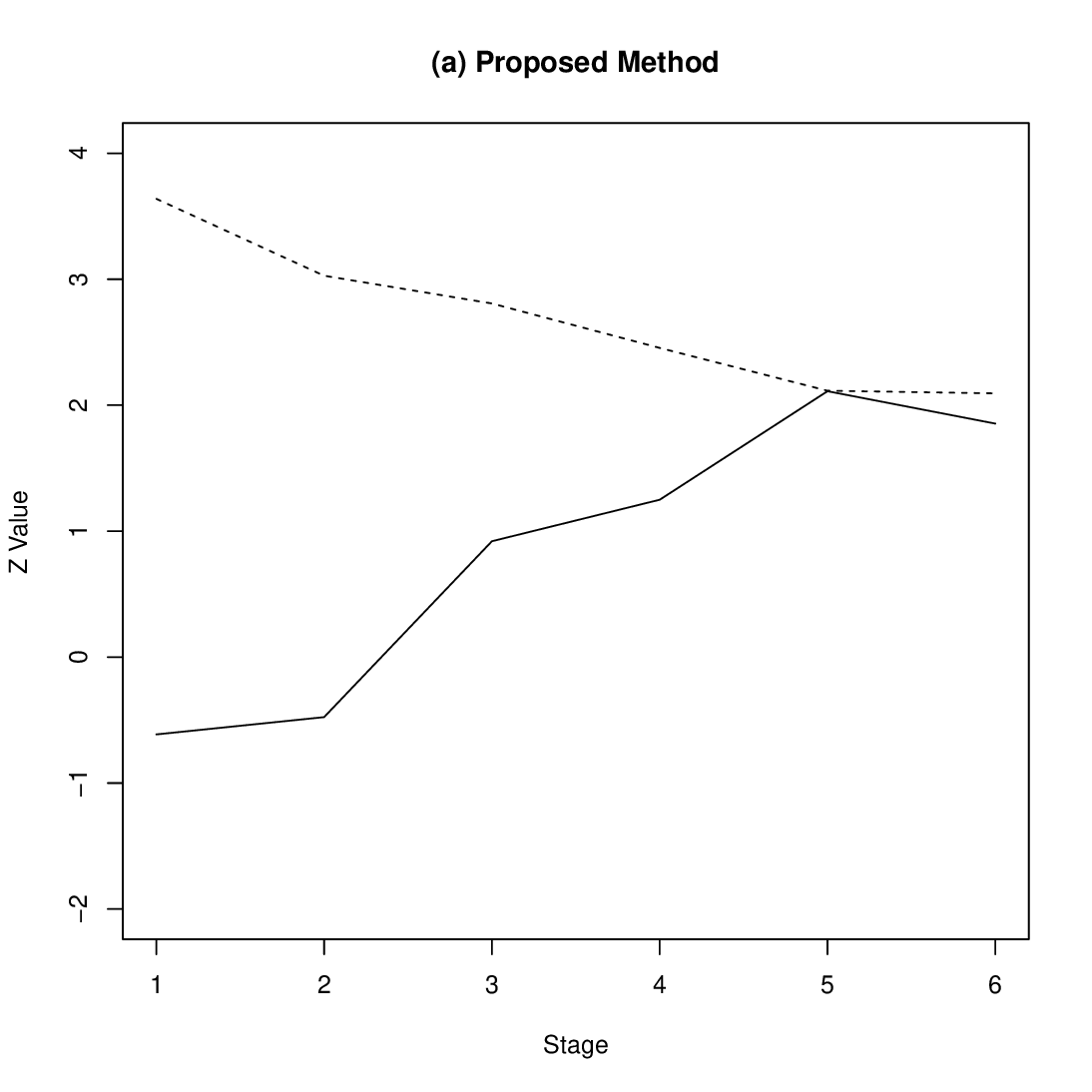}
  \includegraphics[width=8cm]{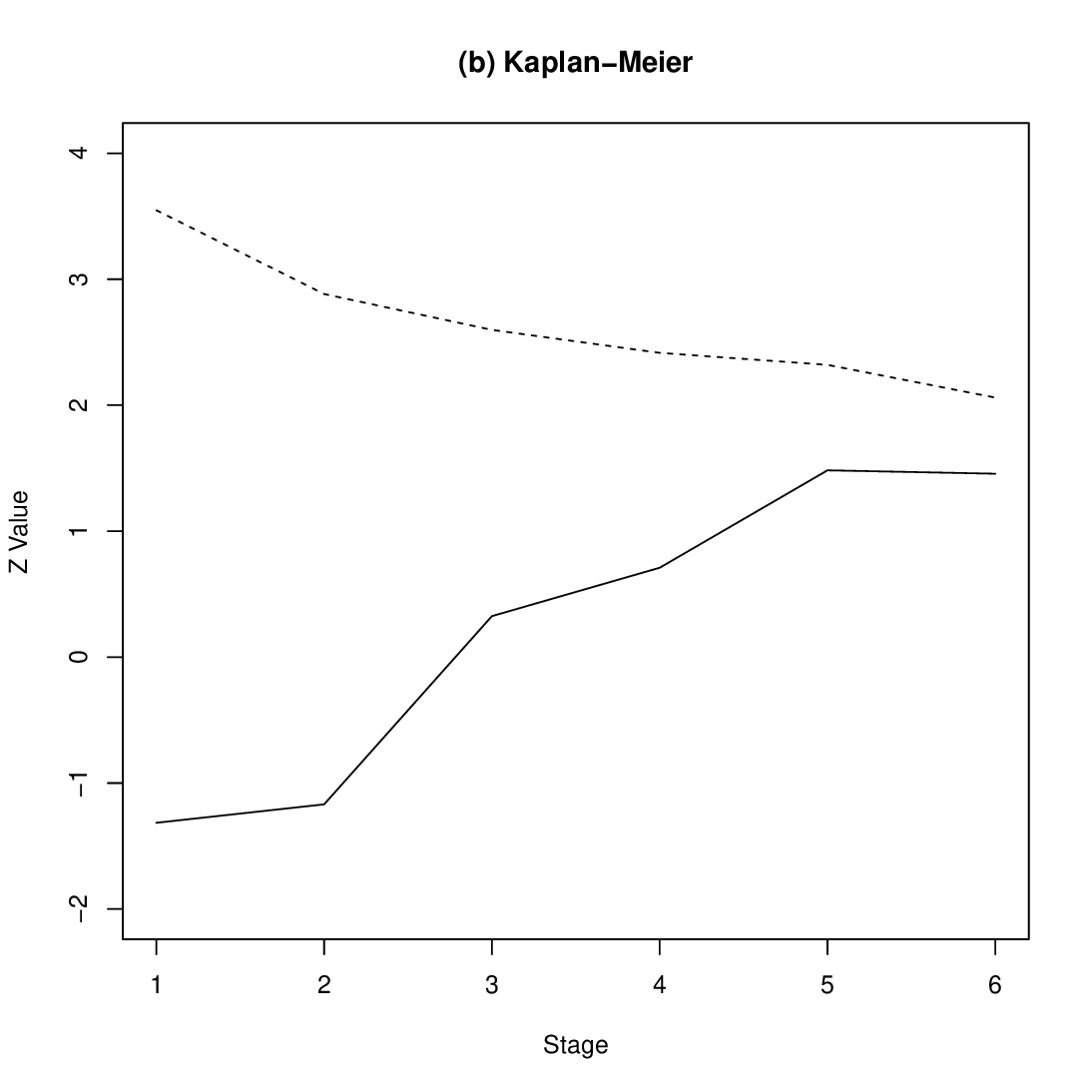}
  \vfill
%   { \footnotesize A - Adjusted Survival Probability, 
%  K - Kaplan-Meier}
 % \includegraphics[width=8cm]{error3.pdf}
 %  {\tiny Sample Size}%
%Figure 1b
\end{figure}

For PFS, standardized test statistics for the group sequential test using the proposed adjusted survival probability method are shown in Figure 3a;
the group sequential Kaplan-Meier based test  estimates are shown in Figure 3b. % 
Both methods found no significant difference at any stage.
Using a group sequential design with the proposed method yielded 
consistent results with the original group sequential Kaplan-Meier analysis,
but also takes into account 
potentially influential covariates. 

\begin{figure}[h]
 \caption{Group sequential test statistics and critical values, overall survival. Dashed lines indicate critical values; solid lines indicate test statistics. The proposed method's test statistics exceeds their corresponding critical values at stages 4-6, while the Kaplan-Meier based comparison's statistics do not exceed the critical values at any stage.}
  \centering
%\raggedright
  \includegraphics[width=8cm]{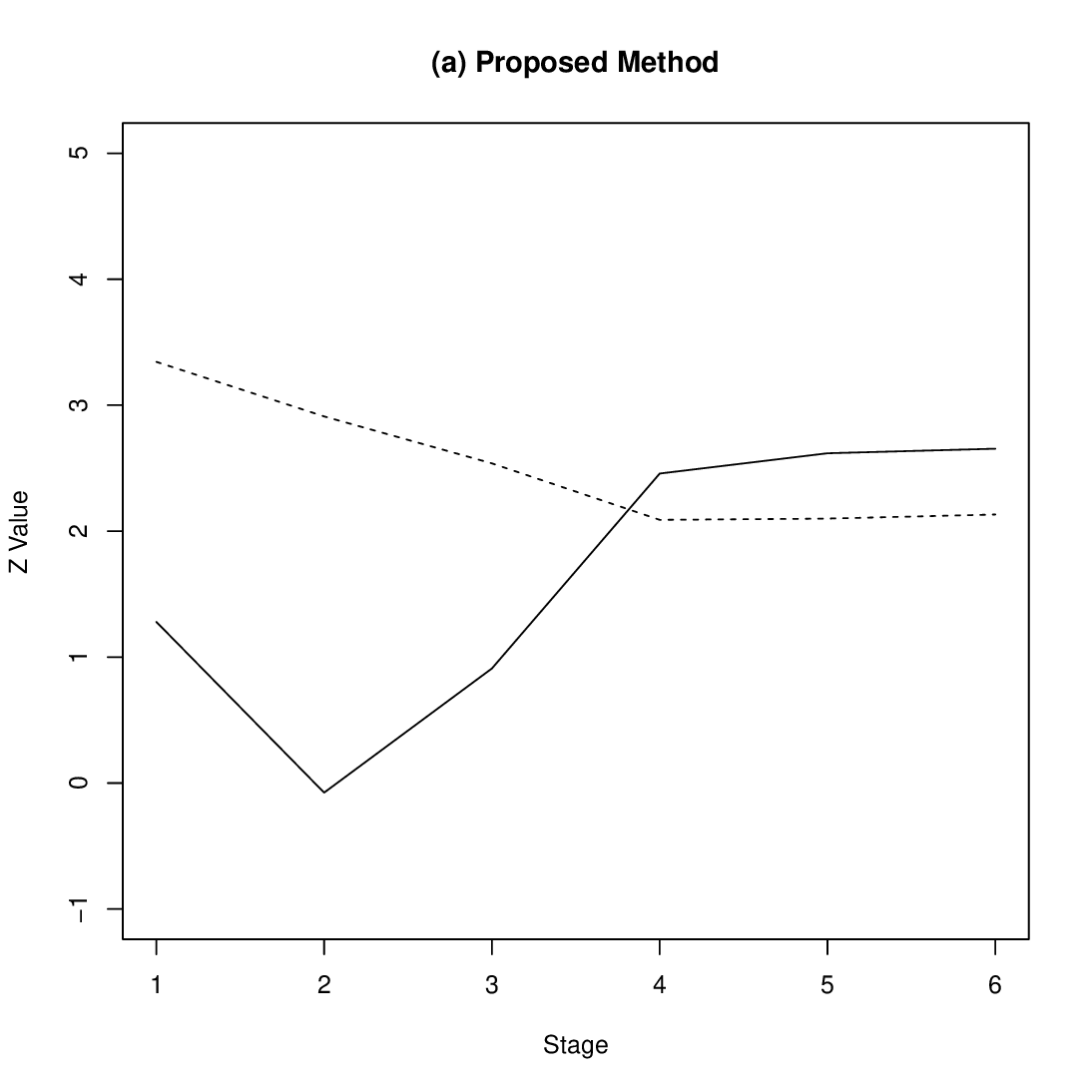}
  \includegraphics[width=8cm]{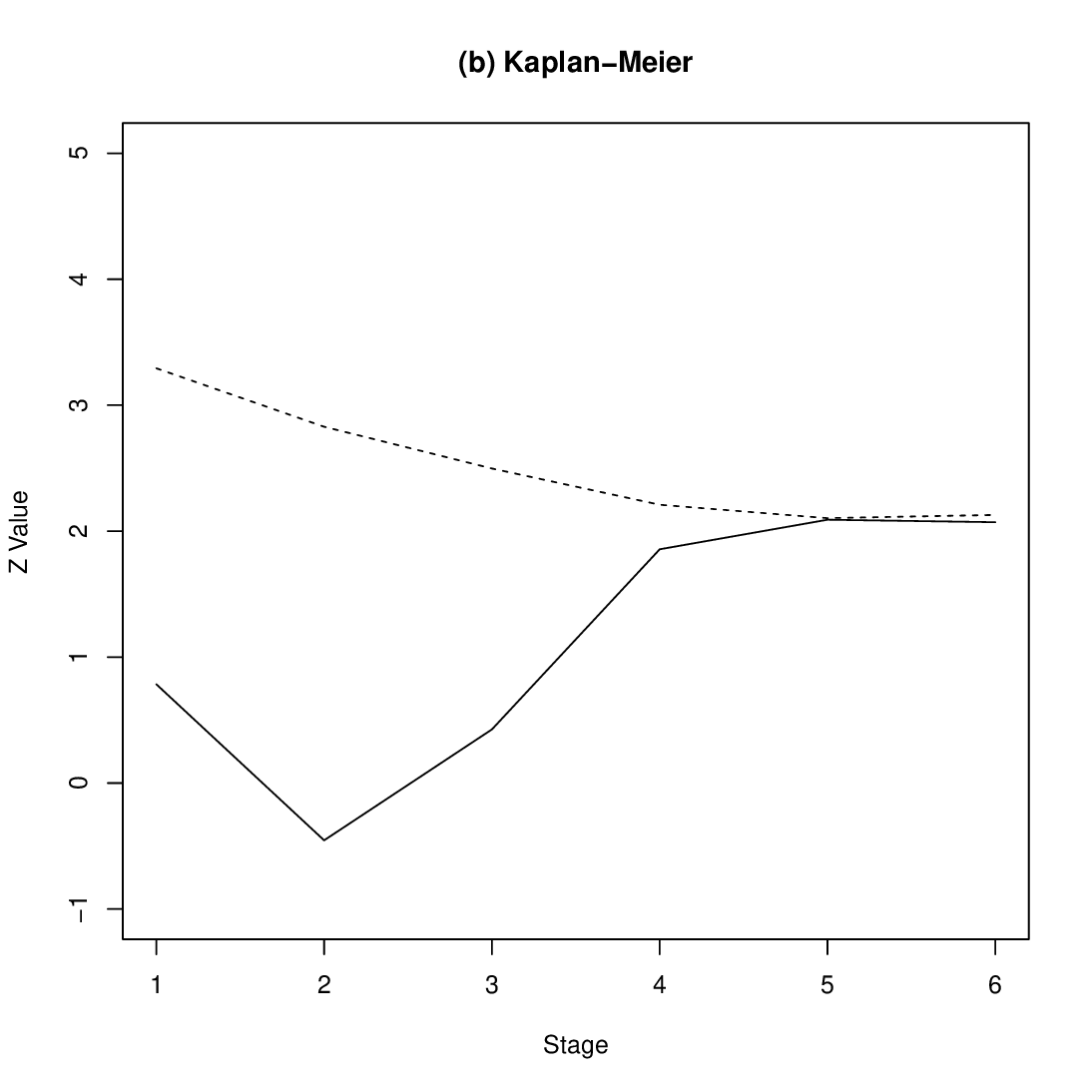}
  \vfill
%   { \footnotesize A - Adjusted Survival Probability, 
%  K - Kaplan-Meier}
 % \includegraphics[width=8cm]{error3.pdf}
 %  {\tiny Sample Size}%
%Figure 1b
\end{figure}

%\begin{figure}[h]
% \caption{Test Statistics vs Critical Values, PFS. Dashed lines indicate critical values, solid lines indicate test statistics}
%  \centering
%\raggedright
%\begin{subfigure}{0.45\textwidth}
 % \includegraphics[width=8cm]{adj-os.eps}
%  \centering
 % \caption{} \label{fig:1a}
%\end{subfigure}
%\hspace*{\fill}
%\begin{subfigure}{0.45\textwidth}

 % \includegraphics[width=8cm]{km-os.eps}
 % \centering
 % \caption{} \label{fig:1b}
%\end{subfigure}
 % \vfill
%   { \footnotesize A - Adjusted Survival Probability, 
%  K - Kaplan-Meier}
 % \includegraphics[width=8cm]{error3.pdf}
 %  {\tiny Sample Size}%
%Figure 1b
%\end{figure}

For OS, the proposed covariate-adjusted GS test statistics %for the GS test using the proposed adjusted survival probability method 
are shown in Figure 4a, %were $ Z_{a1} =1.280, Z_{a2} =-0.076, Z_{a3} = 0.910,  Z_{a4} =2.457,  Z_{a5} = 2.618,Z_{a6} =  2.654$,
whereas the Kaplan-Meier based GS test statistics are shown in Figure 4b. % were $Z_{k1} = 0.784, Z_{k2} = -0.454,  Z_{k3} =0.426,  Z_{k4} =1.856,  Z_{k5} =2.091,  Z_{k6} =2.071$.
%For the Kaplan-Meier analysis,
% critical values were  $c_{k1} = 3.293, c_{k2} =2.829, c_{k3} =2.497, c_{k4} =2.209, c_{k5} = 2.103,  c_{k6} =2.123$;
 The Kaplan-Meier GS test found no significant difference in survival probabilities between the two groups at any stage. 
For the proposed method, %interim critical values were $ c_{a1} =3.344, c_{a2} =2.911, c_{a3} =2.538, c_{a4} =2.090, c_{a5} =2.109, c_{a6} =2.132 $;
a significant difference was found at 6 years, showing higher 2 year survival for the Haplo arm compared to UCB.
By adjusting for covariates and employing a group sequential design
with the proposed method,
a significant difference in overall survival 
would be found 2 years before the original study ended, while a GS Kaplan-Meier analysis would not have found a notable difference at any point during the trial. 

\section{Discussion}
\label{s:discuss}

In this paper, we have developed covariate-adjusted group sequential tests for comparing survival probabilities at a fixed time point under a NPH setting. %Covariate-adjusted estimates of treatment specific survival functions were obtained using a treatment group stratified proportional hazards model.
The proposed covariate-adjusted group sequential test statistics under the treatment stratified PH model enjoy an independent increments structure asymptotically,
%These methods provide test statistics that possess an independent increments structure asymptotically, 
allowing for the application of available methods to determine critical values for testing.
Simulations %were performed to assess the finite sample performance of the proposed tests;
demonstrate that the proposed GS tests met targeted type I error rate and power specifications and were robust to both the PH assumption and covariate effect size. 
A re-analysis of the BMT CTN 1101 trial illustrated the application of these methods.

%The methods in this paper discuss a group sequential, covariate-adjusted test %of the equality
%of survival probabilities at a fixed time point for two treatment arms. 
There are a couple extensions to the methods discussed in this paper. % of interest. % for the methods discussed in this manuscript.
First, it is of interest to investigate methods for sample size and power calculation that use adjusted survival probability methods to assist with designing
clinical trials that will use the adjusted survival probability as an endpoint. 
A future study will examine potential methods for accurate sample size and power
calculations for fixed sample and group sequential studies that use the adjusted survival probability methods.
Approaches using sample size formulas and simulations will be explored for
generating sample size/power estimates and will be compared using empirical studies.
Second, we are interested in investigating GS testing of another endpoint,
the restricted mean survival time (RMST), which is useful in NPH settings. 
The RMST is defined as the expected time survived over a specified interval of time and can be shown to be equal to the area under the survival curve over this interval. 
When comparing RMSTs between treatment groups, the treatment effect has a clear and clinically useful interpretation as the amount of lifetime saved over an interval for one group compared to another. The RMST is an alternative to the hazard ratio in the design and analyses of clinical trials with time to event outcomes, 
and has been studied in fixed sample designs \citep{RoystParm2013}. GS tests for comparing unadjusted RMSTs between treatments have been investigated \citep{MurrTsia99}, but covariate adjustment has yet to be incorporated for restricted mean comparisons in the sequential design setting. 
A group sequential, covariate-adjusted test of the equality of RMSTs will be studied in future work. 

\section*{Acknowledgements}
The authors are grateful to the Blood and Marrow Transplant Clinical Trials Network (BMT CTN) for permission to use and re-analyze the 1101 trial data. Support for the BMT CTN 1101 trial was provided by grants U10HL069294 and U24HL138660 to the BMT CTN from the National Heart, Lung, and Blood Institute (NHLBI) and the National Cancer Institute (NCI). The Center for International Blood and Marrow Transplant Research (CIBMTR) is supported primarily by NCI, NHLBI, and the National Institute of Allergy and Infectious Diseases grant 5U24CA076518; and from HHSH234200637015C (HRSA/DHHS) to the Center for International Blood and Marrow Transplant Research; NHLBI and NCI grant 4U10HL069294; the Health Resources and Services Administration, Department of Health and Human Services contract HHSH250201200016C; and the Office of Naval Research grants N00014-17-1-2388 and N00014-16-1-2020. The content is solely the responsibility of the authors and does not necessarily represent the official views of the aformentioned parties.

\section*{Data Availability Statement}
The data used for the examples in Section 4 are available
from the Blood and Marrow Transplant Clinical Trials Network. Restrictions apply to the availability
of these data, which were permitted to be analyzed by the
authors in this paper. Data are available from the authors with the permission of the Blood and Marrow Transplant Clinical Trials Network.

%\section*{Supporting Information}
%Details on deriving Theorem 1 in Section 2 are available with this paper on arXiv. 

\section*{Supplementary Materials}
\section{Proof of Theorem 1}
\setcounter{equation}{3}

\noindent
{\bf Regularity conditions}
\smallskip

Throughout this paper,
let $\bfbeta_0$ denote the true value of $\bfbeta$ under model (1). For a vector $\bfy$, write $||\bfy|| = \max_i |({\bfy})_i|$ and $|{\bf y}| = \sqrt {\sum y_i^2}$.
For a matrix $\bfY$, write $||{\bfY}|| = \max_{i, j} |({\bfY})_{i, j}|$.
We assume the following regularity conditions
under the treatment-stratified proportional hazards regression model (1),
which are similar to those of Fleming and Harrington (1991) under the proportional hazards regression model.

\begin{enumerate}[label=(\Alph*)]
\item
The survival time $\tau$ is such that $\int_0^{\tau} \lambda_{0i} (t) dt < \infty$ for $i=0,1$.

\item
 For $S_{0i}(\bfbeta,u,t)$, $S_{1i}(\bfbeta,u,t)$, and $S_{2i}(\bfbeta,u,t)$,
%defined in (7) and (14) with $i=0,1$,
there exists a neighborhood ${\cal B}$ of $\bfbeta_0$ and, respectively, scalar, vector, and matrix functions
$s_{0i}(\bfbeta,u,t)$, $s_{1i}(\bfbeta,u,t)$, and $s_{2i}(\bfbeta,u,t)$
defined on ${\cal B} \times [0,\tau] \times [0, \tau]$ such that, for $i = 0, 1$ and $k=0, 1, 2$,
$$\sup_{u,t \in [0, \tau], \bfbeta \in {\cal B}} \bigl|\bigl|S_{ki}(\bfbeta,u,t)-s_{ki}(\bfbeta,u,t)\bigl|\bigl|
\parrow 0, \qquad {\rm as}~n \to \infty.$$
Write
$$e_i(\bfbeta,u,t) = {s_{1i}(\bfbeta,u,t) \over s_{0i}(\bfbeta,u,t)}, \qquad
v_i(\bfbeta,u,t) = {s_{2i}(\bfbeta,u,t) \over s_{0i}(\bfbeta,u,t)} - e_i(\bfbeta,u,t) e_i^T(\bfbeta,u,t).$$
Then, for all $\bfbeta \in {\cal B}$ and $0 \leq u,t \leq \tau$,
$${\partial s_{0i} (\bfbeta,u,t) \over \partial \bfbeta} = s_{1i}(\bfbeta,u,t), \qquad
{\partial^2 s_{0i}(\bfbeta,u,t) \over \partial\bfbeta\partial\bfbeta^T} = s_{2i}(\bfbeta,u,t), \qquad i=0,1.$$

\item
There exists a $\delta > 0$ such that, for $i = 0, 1$,
$${1 \over \sqrt{n_i}} \sup_{1 \leq j \leq n_i, 0 \leq u,t \leq \tau} \bigl|\bfZ_{ij}\bigl| Y_{ij}(u,t)
I(\bfbeta_0^T \bfZ_{ij} > - \delta |\bfZ_{ij}|) \parrow 0, \qquad {\rm as}~n \to \infty.$$

\item
 For $i=0,1$ and $k = 0, 1, 2$,
the functions $s_{ki}(\bfbeta,u,t)$ are bounded and $s_{0i}(\bfbeta,u,t)$ is bounded away from 0 on
${\cal B} \times [0, \tau] \times [0, \tau]$.
The family of functions $s_{ki}(\cdot,u,t)$, $0 \leq u,t \leq \tau$, is an equicontinuous family at $\bfbeta_0$.
See Definition 8.2.1 of Fleming and Harrington (1991, page 289) for the definition
of an equicontinuous family of real-valued functions.

\item
For each $i = 0,1$, there exists a constant $\rho_i \in (0, 1)$ such that, as $n \to \infty$,
\Eq{align}
{
{n_i \over n} \longrightarrow \rho_i.
}

\item
There exists $\tau_0 \in [0,\tau]$ such that
the matrix $\Sigma(\bfbeta_0,u,t)$ is positive definite for all $u,t \in [\tau_0,\tau]$, where
\Eq{align}
{
\Sigma(\bfbeta,u,t)
=\sum_{i=0}^1 \rho_i \int_0^t v_i(\bfbeta,u,s)s_{0i}(\bfbeta,u,s) \lambda_{0i}(s)ds.
}
Write $\Sigma(\bfbeta,u)=\Sigma(\bfbeta,u,u)$.

\item
 For each $i = 0,1$, there exists a function $\pi_i(u,t)$ with $\pi_i(\tau,\tau)>0$ such that
$$\sup_{0 \leq u,t < \tau} \biggl|{1 \over n_i}\sum_{j=1}^{n_i}Y_{ij}(u,t) - \pi_i(u,t)\biggl| \parrow 0.$$
as $n \to \infty$.
\end{enumerate}

We remark that the quantities $e_i(\bfbeta,u,t)$ and $v_i(\bfbeta,u,t)$ defined in Condition (B)
actually do not depend on the calendar time $u$ under model (1)
because $E_{ij}$ and $(T_{ij}, C_{ij}, \bfZ_{ij})$ are independent.
Consequently, we will write throughout that 
$e_i(\bfbeta,u,t)=e_i(\bfbeta,t)$ and $v_i(\bfbeta,u,t)=v_i(\bfbeta,t)$ for $i=0,1$.

\noindent
{\bf Incremental counting processes}
\smallskip

According to Tsiatis, Boucher, and Kim (1995),
we define a right continuous filtration $\{\cF_t: 0 \leq t \leq \tau\}$
as the increasing sequence of sigma algebras generated by
$$\mathcal{F}_t = \sigma\bigl\{E_{ij}, \bfZ_{ij}, I\bigl(X_{ij}^* \leq s, \delta_{ij}^*= 1\bigr),
I\bigl(X_{ij}^* \leq s, \delta_{ij}^*= 0\bigr):  s \leq t, i=0,1, j=1, \ldots, n_i\bigr\},$$
where $X_{ij}^* = \min\{T_{ij},C_{ij}\}$ and $\delta_{ij}^* = I(T_{ij} \leq C_{ij})$.
For each fixed calendar time $u$,
it follows from Theorem 4.2.1 of Fleming and Harrington (1991, page 131) that under model (1),
$M_{ij}(\bfbeta,u,t)=N_{ij}(u,t)-A_{ij}(\bfbeta,u,t)$
is a locally square-integrable martingale with respect to $\mathcal{F}_t$ with predictable variation process
$\langle M_{ij}(\bfbeta,u, \cdot), M_{ij}(\bfbeta,u,\cdot)\rangle (t) = A_{ij}(\bfbeta,u,t)$,
where $A_{ij}(\bfbeta,u,t) = \int_0^t Y_{ij}(u,s) \exp(\bfbeta^T\bfZ_{ij}) \lambda_{0i}(s)ds$
is the compensator process of the counting process $N_{ij}(u,t)$
with respect to $\{{\mathcal F}_t: ~ 0 \leq t \leq \tau\}$ at each calendar time $u$.

Using the same device as in Tsiatis et al. (1995) and Jennison and Turnbull (1997),
we define $\cF_t$-adapted incremental counting processes associated with individual $j$'s observation in group $i$
between successive pairs of interim analyses, i.e.
%$$DN_{ij}(u_1,t) = N_{ij}(u_1,t), \quad DN_{ij}(u_k,t)=N_{ij}(u_k,t)-N_{ij}(u_{k-1},t), \quad k=2, \ldots, K.$$
$DN_{ij}(u_1,t) = N_{ij}(u_1,t)$ and $DN_{ij}(u_k,t)=N_{ij}(u_k,t)-N_{ij}(u_{k-1},t)$ for $k=2, \ldots, K.$
%Under the treatment-stratified proportional hazards model (2.1),
Under model (1),
the compensator process of the counting process $DN_{ij}(u_k,t)$
with respect to $\{{\cal F}_t: ~ 0 \leq t \leq \tau\}$ is given by
$DA_{ij}(u_1,t) = A_{ij}(u_1,t)$ and
$DA_{ij}(\bfbeta,u_k,t) = A_{ij}(\bfbeta,u_k,t)-A_{ij}(\bfbeta,u_{k-1},t)
= \int_0^t \lambda_{0i}(s) \exp({\boldsymbol\beta}^T\bfZ_{ij}) DY_{ij}(u_k,s)ds$ for $k=2, \ldots, K$,
where $DY_{ij}(u_1,t) = Y_{ij}(u_1,t)$ and  $DY_{ij}(u_k,t) = Y_{ij}(u_k,t) - Y_{ij}(u_{k-1},t)$ for $k=2, \ldots, K$.
As a result, the stochastic process
%$$DM_{ij}(\bfbeta,u_k,t) = M_{ij}(\bfbeta,u_k,t)-M_{ij}(\bfbeta,u_{k-1},t)
%%= \{N_{ij}(u_k,t)-A_{ij}(\bfbeta,u_k,t)\} - \{N_{ij}(u_{k-1},t)-A_{ij}(\bfbeta,u_{k-1},t)\} \cr
%%&\quad= \{N_{ij}(u_k,t)-N_{ij}(u_{k-1},t)\} - \{A_{ij}(\bfbeta,u_k,t)-A_{ij}(\bfbeta,u_{k-1},t)\}
%= DN_{ij}(\bfbeta,u_k,t) - DA_{ij}(\bfbeta,u_k,t)$$
$DM_{ij}(\bfbeta,u_k,t) = M_{ij}(\bfbeta,u_k,t)-M_{ij}(\bfbeta,u_{k-1},t)
= DN_{ij}(\bfbeta,u_k,t) - DA_{ij}(\bfbeta,u_k,t)$
is a local square-integrable martingale with respect to $\cF_t$ for $k=1, \ldots, K$, where $M_{ij}(\bfbeta,u_0,t)=0$.
Write $DA_{ij}(u_k,t) = DA_{ij}(\bfbeta_0,u_k,t)$ and $DM_{ij}(u_k,t) = DM_{ij}(\bfbeta_0,u_k,t)$.

Notice that
(i) the martingales $DM_{i_1j_1}(\bfbeta,u_{k_1},t)$ and $DM_{i_2j_2}(\bfbeta,u_{k_2},t)$ are orthogonal 
if $i_1 \not= i_2$ or $j_1 \not= j_2$, because individuals are independent in the pooled sample; 
and (ii) the martingales $DM_{ij}(\bfbeta,u_k,t)$ and $DM_{ij}(\bfbeta,u_l,t)$ are orthogonal if $k \not= l$,
because they cannot jump at the same time.
%This orthogonality is essential in establishing the asymptotic joint normality of the random vector
%$\bigl\{\sqrt{n}\{\hat S_1(u_k,t_0)-\hat S_0(u_k,t_0)\},~ k=1, \ldots, K\bigr\}$.
In addition,
the counting process $\{DN_{ij}(u_k,t): 0 \leq t \leq \tau,~i=0,1,~ j=1, \ldots, n_i,~k=1, \ldots, K\}$
is a multivariate counting process,
and the compensator $DA_{ij}(u_k,t)$ of $DN_{ij}(u_k,t)$ is continuous.

\noindent
{\bf Finite dimensional distributions of $\sqrt{n} \{\hat S_1(u,t_0)-\hat S_0(u,t_0)\}$}
\smallskip

Under the null hypothesis $H_0: S_0(t_0)=S_1(t_0)$,
based on the asymptotic expansion of the average survival curve for each treatment group from Zucker (1998, page 704),
we can write at calendar time $u$, as $n \to \infty$,
\Eq{align}
{
\label{eq:SP_asymp}
&\sqrt{n}\{\hat S_1(u,t_0) - \hat S_0(u,t_0)\}
= \sqrt{n} \{\hat S_1(u,t_0) -  S_1(t_0)\} - \sqrt{n} \{\hat S_0(u,t_0) - S_0(t_0)\} \cr
&= c_{01}(t) C_0(\bfbeta_0,u,t_0)-c_{11}(t) C_1(\bfbeta_0,u,t_0)
+ D^T(\bfbeta_0,t_0) \sqrt{n}\{\hat\bfbeta(u)-\bfbeta_0\} \cr
&= c_{01}(t) C_0(\bfbeta_0,u,t_0)-c_{11}(t) C_1(\bfbeta_0,u,t_0)
+ D^T(\bfbeta_0,t_0) \Sigma^{-1}(\bfbeta_0,u) {1 \over \sqrt n}U(\bfbeta_0,u) + o_p(1).
}
To establish the asymptotic joint normality of 
$\bigl\{\sqrt{n}\{\hat S_1(u_k,t_0)-\hat S_0(u_k,t_0)\}, ~ k=1, \ldots, K\bigr\}$
under $H_0: S_0(t_0)=S_1(t_0)$ at the calendar times $0< t_0 \leq u_1 < u_2 < \cdots < u_K $,
it follows from the asymptotic expansion (\ref{eq:SP_asymp}) that
we can consider the joint distribution of the $(p+2)K$-dimensional random vectors 
$\{C_0(\bfbeta_0,u_k,t): ~ k=1, \ldots, K\}$, $\{C_1(\bfbeta_0,u_k,t):~ k=1, \ldots, K\}$, and 
$\{n^{-1/2}U(\bfbeta_0,u_k):~ k=1, \ldots, K\}$
by expressing them as stochastic integrals with respect to the martingales 
$\{DM_{ij}(u_m,t):$ $i=0,1, j=1, \ldots, n_i, m=1, \ldots, K\}$, i.e.
\Eq{align*}
{
& U(\bfbeta_0,u_k,t) = 
 \sum_{m=1}^k \sum_{i=0}^1\sum_{j=1}^{n_i} \int_0^t \bigl\{\bfZ_{ij} - E_i(\bfbeta_0,u_k,s)\bigr\}
DM_{ij}(u_m,ds), \cr
& C_i(\bfbeta_0,u_k,t) 
= {\sqrt{n} \over n_i} \sum_{m=1}^k \sum_{j=1}^{n_i}
\int_0^t {1 \over S_{0i}(\bfbeta_0,u_k,s)} I\biggl(\sum_{j=1}^{n_i}Y_{ij}(u_k,s) > 0\biggr)
DM_{ij}(u_m,ds), \quad i=0,1.
}
Considered as processes in $t$, it is seen from these expressions that
$n^{-1/2}U(\bfbeta_0,u_k)$ and $C_i(\bfbeta_0,u_k,t)$
are linear combinations of stochastic integrals of
predictable and locally bounded processes with respect to the local square-integrable martingales $DM_{ij}(u_m,t)$
for $i=0,1$, $j=1, \ldots, n_i$, $m=1, \ldots, K$,
and hence are also local square-integrable martingales with mean zero.

According to Theorem 8.2.1 of Fleming and Harrington (1991, page 290),
the partial likelihood score vectors $\{n^{-1/2}U(\bfbeta_0,u_k):~ k=1, \ldots, K\}$
satisfy the Lindeberg condition (3.18) of Fleming and Harrington (1991, page 228).
To show that the random vectors $\{C_i(\bfbeta_0,u_k,t):~ k=1, \ldots, K\}$ satisfy the Lindeberg condition 
for $t \in [0, \tau]$, write
\Eq{align*}
{
& C_{in, \varepsilon}(\bfbeta_0,u_k,t)
= \sum_{m=1}^k \sum_{j=1}^{n_i} \int_0^t  H_{in}(\bfbeta_0,u_k,s)
I(|H_{in}(\bfbeta_0,u_k,s)| \geq \varepsilon) d DM_{ij}(u_m,s), \qquad i=0,1,
}
where $Y_i(u,t)=\sum_{j=1}^{n_i} Y_{ij}(u,t)$ is the number at risk at survival time $t$ for $i=0,1$ and
$$H_{in}(\bfbeta_0,u,t) = {\sqrt{n}I(Y_i(u,t) > 0) \over n_iS_{0i}(\bfbeta_0,u,t)}, \qquad i=0,1.$$
Using Theorem 2.4.3 of Fleming and Harrington (1991, page 70) along with the linearity of $\langle\cdot, \cdot\rangle$,
we have
\Eq{align*}
{
& \bigl\langle C_{in, \varepsilon}(\bfbeta_0,u_k,\cdot), C_{in, \varepsilon} (\bfbeta_0,u_k,\cdot)\bigr\rangle (t) \cr
&\quad = \sum_{m=1}^k \sum_{j=1}^{n_i} \int_0^t
H_{in}^2(\bfbeta_0,u_k,s) I(|H_{in}(\bfbeta_0,u_k,s)| \geq \varepsilon)
\lambda_{0i}(s) \exp({\boldsymbol\beta}_0^T\bfZ_{ij}) DY_{ij}(u_m,s)ds \cr
&\quad = \sum_{j=1}^{n_i} \int_0^t H_{in}^2(\bfbeta_0,u_k,s) I(|H_{in}(\bfbeta_0,u_k,s)| \geq \varepsilon)
\lambda_{0i}(s) \exp({\boldsymbol\beta}_0^T\bfZ_{ij}) Y_{ij}(u_k,s)ds \cr
&\quad = \int_0^t \sqrt{n} H_{in}(\bfbeta_0,u_k,s) I(|H_{in}(\bfbeta_0,u_k,s)| \geq \varepsilon) d\Lambda_{0i}(s),
}
where $H_{in}^2(\bfbeta_0,u_k,s) n_iS_{0i}(\bfbeta_0,u_k,s) = \sqrt{n} H_{in}(\bfbeta_0,u_k,s)$.
Condition (G) implies that \\
$P\bigl\{\sup_{0 \leq s \leq t} |I(Y_i(u_k,s) > 0)-1| > \varepsilon\bigr\}
= \{1-\pi_i(u_k,t-)\}^{n_i} \to 0$ as $n \to \infty$ for any $\varepsilon>0$,
and hence 
$\sup_{0 \leq s \leq t} |I(Y_i(u_k,s) > 0)-1| \parrow 0$.
Furthermore, since $s_{0i}(\bfbeta_0,u_k,s)$ is bounded away from 0 for $s \in [0,\tau]$ by Condition (D),
it follows from Conditions (B) and (E) that
\Eq{align*}
{
& \sup_{0 \leq s \leq t} \bigl|\sqrt{n} H_{in}(\bfbeta_0,u_k,s)\bigl| 
\leq {n \over n_i} \sup_{0 \leq s \leq t} {1 \over S_{0i}(\bfbeta_0,u_k,s)}
\leq {n \over n_i} {1 \over S_{0i}(\bfbeta_0,u_k,t)}
\parrow {1 \over \rho_i s_{0i}(\bfbeta_0,u_k,t)}.
}
This implies that
$\sup_{0 \leq s \leq t} \bigl|H_{in}(\bfbeta_0,u_k,s)\bigl| \parrow 0$,
which further implies that  $\sup_{0 \leq s \leq t} I(|H_{in}(\bfbeta_0,u_k,s)| \geq \varepsilon) \parrow 0$.
Combining these results yields
\Eq{align*}
{
& \bigl\langle C_{in, \varepsilon}(\bfbeta_0,u_k,\cdot), C_{in, \varepsilon} (\bfbeta_0,u_k,\cdot)\bigr\rangle (t)
= \int_0^t \sqrt{n} H_{in}(\bfbeta_0,u_k,s) I(|H_{in}(\bfbeta_0,u_k,s)| \geq \varepsilon) d\Lambda_{0i}(s) \cr
&\quad \leq \sup_{0 \leq s \leq t} \biggl\{\sqrt{n}H_{in}(\bfbeta_0,u_k,s) \biggr\}
\sup_{0 \leq s \leq t}\biggl\{I(|H_{in}(\bfbeta_0,u_k,s)| \geq \varepsilon)\biggr\} \Lambda_{0i}(t) \parrow 0.
}
Setting $t=t_0$, we have proved that
the random vectors
$\{C_0(\bfbeta_0,u_k,t_0):~ k=1, \ldots, K\}$, $\{C_1(\bfbeta_0,u_k,t_0): k=1, \ldots, K\}$, and
$\{n^{-1/2}U(\bfbeta_0,u_k):~ k=1, \ldots, K\}$
satisfy the Lindeberg condition (3.18) of Fleming and Harrington (1991, page 228).

Consequently, an application of Theorem 5.3.5 of Fleming and Harrington (1991, page 227)
yields the asymptotic multivariate normal distribution of the $(p+2)K$-dimensional random vectors
$\{C_0(\bfbeta_0,u_k,t_0):~ k=1, \ldots, K\}$, $\{C_1(\bfbeta_0,u_k,t_0):~ k=1, \ldots, K\}$,
and $\{n^{-1/2}U(\bfbeta_0,u_k):~ k=1, \ldots, K\}$
by Rebolledo's martingale central limit theorem.
In particular, the $(p+2)K$-dimensional random vector
\be
\label{eq:comp_vector}
\biggl\{ C_0(\bfbeta_0,u_k,t_0),~ C_0(\bfbeta_0,u_k,t_0),~
{1 \over \sqrt{n}} U(\bfbeta_0,u_k): ~~ k=1, \ldots, K\biggr\}
\ee
converges in distribution to a multivariate normal random vector with mean zero. 

Since
$\bigl(\sqrt{n}\{\hat S_1(u_1,t_0)-\hat S_0(u_1,t_0)\}, \ldots, 
\sqrt{n}\{\hat S_1(u_K,t_0)-\hat S_0(u_K,t_0)\}\bigr)$
is, by (\ref{eq:SP_asymp}), a linear combination of the components of the $(p+2)K$-dimensional random vector in (\ref{eq:comp_vector}),
it follows from its asymptotic joint normality that the joint distribution of the $K$-dimensional vector statistic
$\bigl(\sqrt{n}\{\hat S_1(u_1,t_0)-\hat S_0(u_1,t_0)\}, \ldots, 
\sqrt{n}\{\hat S_1(u_K,t_0)-\hat S_0(u_K,t_0)\}\bigr)$
is asymptotically multivariate normal with  mean zero under the null hypothesis $H_0: S_0(t_0) = S_1(t_0)$.

To calculate the asymptotic variances and covariances of
$\bigl\{\sqrt{n}\{\hat S_1(u_k,t_0)-\hat S_0(u_k,t_0)\}
- \sqrt{n}\{S_1(t_0)-S_0(t_0)\},~ k=1, \ldots, K\bigr\}$,
note first that
$\bigl\langle C_i(\bfbeta_0,u_k,t_0), U_q(\bfbeta_0,u_l)\bigr\rangle \parrow 0$
for $i=0,1$, $q=1, \ldots, p$, and $1 \leq k, l \leq K$ after some lengthy calculations,
where $U_q(\bfbeta_0,u)$ is the $q$th component of $U(\bfbeta_0,u)$ for $q=1,\ldots, p$.
As a result,
$C_i(\bfbeta_0,u_k,t_0)$ and $U(\bfbeta_0,u_l)$ are asymptotically independent for $i=0,1$ and $1 \leq k, l \leq K$.
Furthermore, since
\Eq{align*}
{
& \langle C_i(\bfbeta_0,u_k,\cdot), C_i(\bfbeta_0,u_l,\cdot) \rangle (t) \cr
&= \biggl\langle
\sum_{m=1}^k\sum_{j=1}^{n_i} {\sqrt{n} \over n_i} 
\int_0^t{I(Y_i(u_k,s)>0) \over S_{0i}(\bfbeta_0,u_k,s)} DM_{ij}(u_m,ds),
\sum_{m=1}^l\sum_{j=1}^{n_i} {\sqrt{n} \over n_i}
\int_0^t{I(Y_i(u_l,s)>0) \over S_{0i}(\bfbeta_0,u_l,s)} DM_{ij}(u_m,ds) \biggr\rangle \cr
&= \sum_{m=1}^k\sum_{j=1}^{n_i}
\biggl\langle \int_0^t {\sqrt{n} \over n_i} {I(Y_i(u_k,s)>0) \over S_{0i}(\bfbeta_0,u_k,s)} DM_{ij}(u_{m},ds),~
\int_0^t {\sqrt{n} \over n_i}{I(Y_i(u_l,s)>0) \over S_{0i}(\bfbeta_0,u_l,s)} DM_{ij}(u_m,ds) \biggr\rangle \cr
&= \sum_{m=1}^k\sum_{j=1}^{n_i} \int_0^t
{n \over n_i^2} {I(Y_i(u_k,s)>0) \over S_{0i}(\bfbeta_0,u_k,s)}
{I(Y_i(u_l,s)>0) \over S_{0i}(\bfbeta_0,u_l,s)}
\lambda_{0i}(s) \exp({\boldsymbol\beta}_0^T\bfZ_{ij}) DY_{ij}(u_m,s)ds \cr
&= \sum_{j=1}^{n_i} \int_0^t {n \over n_i^2} {I(Y_i(u_k,s)>0) \over S_{0i}(\bfbeta_0,u_k,s)}
{I(Y_i(u_l,s)>0) \over S_{0i}(\bfbeta_0,u_l,s)}
\lambda_{0i}(s) \exp({\boldsymbol\bfbeta}_0^T\bfZ_{ij}) Y_{ij}(u_k,s)ds \cr
&= \int_0^t {n \over n_i} {I(Y_i(u_k,s)>0)I(Y_i(u_l,s)>0) \over S_{0i}(\bfbeta_0,u_l,s)}\lambda_{0i}(s)ds
\parrow 
{\gamma_i(\bfbeta_0,u_l,t) \over \rho_i}, \qquad i=0,1,
}
for $1 \leq k \leq l \leq K$,
this implies that the asymptotic covariance of $C_i(\bfbeta_0,u_k,t_0)$ and $C_i(\bfbeta_0,u_l,t_0)$ is
$$\ACov\bigl\{C_i(\bfbeta_0,u_k,t_0), C_i(\bfbeta_0,u_l,t_0)\bigr\}
= {\gamma_i(\bfbeta_0,u_l,t_0) \over \rho_i}, \qquad i=0,1.$$
In particular, the asymptotic variance of $C_i(\bfbeta_0,u_k,t_0)$ is given by
$$\AVar\bigl\{C_i(\bfbeta_0,u_k,t_0)\bigr\} = {\gamma_i(\bfbeta_0,u_k,t_0) \over \rho_i}, \qquad i=0,1.$$
For $0 \leq k,l \leq K$, $C_0(\bfbeta_0,u_k,t_0)$ and $C_1(\bfbeta_0,u_l,t_0)$ are asymptotically independent
because the two treatment-specific samples are independent.
Therefore,
For $1 \leq k \leq l \leq K$,
the asymptotic covariance of
$\sqrt{n} \{\hat S_1(u_k,t_0) - \hat S_0(u_k,t_0)\}$ and
$\sqrt{n} \{\hat S_1(u_l,t_0) - \hat S_0(u_l,t_0)\}$ is calculated as
\Eq{align*}
{
& \sigma(u_k,u_l,t_0) 
= \ACov\bigl(\sqrt{n} \{\hat S_1(u_k,t_0) - \hat S_0(u_k,t_0)\},~
\sqrt{n} \{\hat S_1(u_l,t_0) - \hat S_0(u_l,t_0)\} \bigr) \cr
&\quad= \ACov\biggl[c_{01}(t_0) C_0(\bfbeta_0,u_k,t_0)-c_{11}(t_0) C_1(\bfbeta_0,u_k,t_0)
+ D^T(\bfbeta_0,t_0) \Sigma^{-1}(\bfbeta_0,u_k) {1 \over \sqrt n}U(\bfbeta_0,u_k), \cr
&\qquad\quad c_{01}(t_0) C_0(\bfbeta_0,u_l,t_0)-c_{11}(t_0) C_1(\bfbeta_0,u_l,t_0)
+ D^T(\bfbeta_0,t_0) \Sigma^{-1}(\bfbeta_0,u_l){1 \over \sqrt n}U(\bfbeta_0,u_l)\biggr] \cr
&\quad= c_{01}^2(t_0) \ACov\bigl\{C_0(\bfbeta_0,u_k,t_0), C_0(\bfbeta_0,u_l,t_0)\}
+ c_{11}^2(t_0) \ACov\bigl\{C_1(\bfbeta_0,u_k,t_0), C_1(\bfbeta_0,u_l,t_0)\} \cr
&\qquad + D^T(\bfbeta_0,t_0)
\ACov\bigl\{\sqrt{n}(\hat\bfbeta_{(k)}-\bfbeta_0), \sqrt{n}(\hat\bfbeta_{(l)}-\bfbeta_0)\bigr\}D(\bfbeta_0,t_0) \cr
&\quad= c_{01}^2(t_0) \ACov\bigl\{C_0(\bfbeta_0,u_k,t_0), C_0(\bfbeta_0,u_l,t_0)\}
+ c_{11}^2(t_0) \ACov\bigl\{C_1(\bfbeta_0,u_k,t_0), C_1(\bfbeta_0,u_l,t_0)\} \cr
&\qquad + D^T(\bfbeta_0,t_0) 
\AVar\bigl\{\sqrt{n}(\hat\bfbeta_{(l)}-\bfbeta_0)\bigr\} D^T(\bfbeta_0,t_0) \cr
&\quad= c_{01}^2(t_0) {\gamma_0(\bfbeta_0,u_l,t_0) \over \rho_0}
+ c_{11}^2(t_0) {\gamma_1(\bfbeta_0,u_l,t_0) \over \rho_1}
+ D^T(\bfbeta_0,t) \Sigma^{-1}(\bfbeta_0,u_l) D(\bfbeta_0,t_0).
}
In particular,
the asymptotic variance of $\sqrt{n} \{\hat S_1(u_k,t_0)-\hat S_0(u_k,t_0)\}$ 
is given by
\Eq{align*}
{
& \sigma^2(u_k,t_0) 
= \AVar\bigl(\sqrt{n} \{\hat S_1(u_k,t_0)-\hat S_0(u_k,t_0)\}\bigr) \cr
&\quad = {1 \over \rho_0} c_{01}^2(t) \gamma_0(\bfbeta_0,u_k,t_0)
+ {1 \over \rho_1} c_{11}^2(t) \gamma_1(\bfbeta_0,u_k,t_0)
+ D^T(\bfbeta_0,t_0) \Sigma(\bfbeta_0,u_k)D(\bfbeta_0,t_0). 
}

\vspace{0.5cm}
\noindent
{\bf Tightness of $\sqrt{n} \{\hat S_1(u,t_0)-\hat S_0(u,t_0)\}$}
\smallskip

\noindent
We have shown
that the finite dimensional distributions of $\sqrt{n} \{\hat S_1(u,t_0)-\hat S_0(u,t_0)\}$ 
converge to those of the finite dimensional distributions of $\xi$ under the null hypothesis $H_0: S_0(t_0)=S_1(t_0)$.
To establish the asymptotic convergence of 
$\sqrt{n} \{\hat S_1(u,t_0)-\hat S_0(u,t_0)\}$ to the Gaussian process $\xi$ in distribution,
we need to verify the tightness of $\sqrt{n} \{\hat S_1(u,t_0)-\hat S_0(u,t_0)\}$
when $S_0(t_0)=S_1(t_0)$.

It follows from Theorem 4.3 of Bilias, Gu, and Ying (1997) that
 as $n \to \infty$,
$$\sqrt{n} \{\hat\Lambda_{0i}(u,t_0) - \Lambda_{0i}(t_0)\}
= C_i(\bfbeta_0,u,t_0) - Q_i^T(\bfbeta_0,t_0) \sqrt{n}\bigl\{\hat\bfbeta(u)-\bfbeta_0\bigr\} + o_p(1),$$
where the remainder term $o_p(1)$ is uniform for $u \in [t_0, \tau]$ under Conditions (B) and (D).
According to Theorems 4.2 and 4.3 of Bilias, Gu, and Ying (1997),
$\sqrt{n}\bigl\{\hat\bfbeta(u)-\bfbeta_0\bigr\}$ and $\sqrt{n} \{\hat\Lambda_{0i}(u,t_0) - \Lambda_{0i}(t_0)\}$
converge in distribution to a vector Gaussian process and a Gaussian process, respectively,
which implies that both $\sqrt{n}\bigl\{\hat\bfbeta(u)-\bfbeta_0\bigr\}$ and
$\sqrt{n} \{\hat\Lambda_{0i}(u,t_0) - \Lambda_{0i}(t_0)\}$ are tight for $u \in [t_0,\tau]$.
Since $C_i(\bfbeta_0,u,t_0)$ is asymptotically a linear combination of
$\sqrt{n}\bigl\{\hat\bfbeta(u)-\bfbeta_0\bigr\}$ and $\sqrt{n} \{\hat\Lambda_{0i}(u,t_0) - \Lambda_{0i}(t_0)\}$,
the stochastic process $C_i(\bfbeta_0,u,t_0)$ is also tight for $u \in [t_0,\tau]$ and $i=0,1$.
It now follows from Theorem B.1.6 of Fleming and Harrington (1991, page 340) that for every $\varepsilon > 0$,
\Eq{align}
{
\label{eq:C_betahat_tight}
& \lim_{\delta \downarrow 0} \limsup_{n \to \infty}
P\biggl(\sup_{|u-v| < \delta} |C_i(\bfbeta_0,u,t_0)-C_i(\bfbeta_0,v,t_0)| > \varepsilon\biggr)=0, \cr
& \lim_{\delta \downarrow 0} \limsup_{n \to \infty}
P\biggl(\sup_{|u-v| < \delta}
\bigl|\sqrt{n}\bigl(\hat\bfbeta(u)-\bfbeta_0\bigr)-\sqrt{n}\bigl(\hat\bfbeta(v)-\bfbeta_0\bigr)\bigl|
> \varepsilon\biggr)=0.
}

For any $u, v \in [t_0,\tau]$ satisfying $|u-v| \leq \delta$ with $\delta>0$, since
\Eq{align*}
{
& \biggl|\bigl[c_{01}(t_0) C_0(\bfbeta_0,u,t_0)-c_{11}(t_0) C_1(\bfbeta_0,u,t)
+D^T(\bfbeta_0,t_0) \sqrt{n}\bigl\{\hat\bfbeta(u)-\bfbeta_0\bigr\}\bigr] \cr
&\qquad - \bigl[c_{01}(t_0) C_0(\bfbeta_0,v,t_0)-c_{11}(t_0) C_1(\bfbeta_0,v,t)
+D^T(\bfbeta_0,t_0) \sqrt{n}\bigl\{\hat\bfbeta(v)-\bfbeta_0\bigr\}\bigr]\biggl| \cr
&\quad \leq \bigl|c_{01}(t_0) \{C_0(\bfbeta_0,u,t_0)-C_0(\bfbeta_0,v,t_0)\}\bigl|
 + \bigl|c_{11}(t_0) \{C_1(\bfbeta_0,u,t)-C_1(\bfbeta_0,v,t)\}\bigl| \cr
&\quad + \bigl|D^T(\bfbeta_0,t_0) \bigl\{\sqrt{n}\bigl(\hat\bfbeta(u)-\bfbeta_0\bigr)
-\sqrt{n}\bigl(\hat\bfbeta(v)-\bfbeta_0\bigr)\bigr\}\bigl| \cr
&\quad \leq |c_{01}(t_0)| \sup_{|u-v| < \delta} |C_0(\bfbeta_0,u,t_0)-C_0(\bfbeta_0,v,t_0)|
+ |c_{11}(t_0)| \sup_{|u-v| < \delta} |C_1(\bfbeta_0,u,t)-C_1(\bfbeta_0,v,t_0)| \cr
&\qquad + |D(\bfbeta_0,t_0)|
\sup_{|u-v| \leq \delta}\bigl|\sqrt{n}\bigl(\hat\bfbeta(u)-\bfbeta_0\bigr)
-\sqrt{n}\bigl(\hat\bfbeta(v)-\bfbeta_0\bigr)\bigl|,
}
it follows from (\ref{eq:C_betahat_tight}) that for every $\varepsilon > 0$,
\Eq{align*}
{
& \lim_{\delta \downarrow 0} \limsup_{n \to \infty} P\biggl(\sup_{|u-v| \leq \delta}
\biggl|\bigl[c_{01}(t_0) C_0(\bfbeta_0,u,t_0)-c_{11}(t_0) C_1(\bfbeta_0,u,t)
+D^T(\bfbeta_0,t_0) \sqrt{n}\bigl\{\hat\bfbeta(u)-\bfbeta_0\bigr\}\bigr] \cr
&\quad - \bigl\{c_{01}(t_0) C_0(\bfbeta_0,v,t_0)-c_{11}(t_0) C_1(\bfbeta_0,v,t)
+D^T(\bfbeta_0,t_0) \sqrt{n}\bigl\{\hat\bfbeta(v)-\bfbeta_0\bigr\}\bigr]\biggl|> \varepsilon \biggr) \cr
&\quad \leq \lim_{\delta \downarrow 0} \limsup_{n \to \infty} P\biggl(
\sup_{u \in [t_0, \tau]} |C_0(\bfbeta_0,u,t_0)-C_0(\bfbeta_0,v,t_0)| > {\varepsilon \over 3|c_{01}(t_0)|} \biggr) \cr
&\qquad + \lim_{\delta \downarrow 0} \limsup_{n \to \infty} 
P\biggl(\sup_{u \in [t_0, \tau]} |C_1(\bfbeta_0,u,t)-C_1(\bfbeta_0,v,t_0)|
> {\varepsilon \over 3|c_{11}(t_0)|}\biggr) \cr
&\qquad + \lim_{\delta \downarrow 0} \limsup_{n \to \infty}
P\biggl(\sup_{|u-v| \leq \delta}\bigl|\sqrt{n}\bigl\{\hat\bfbeta(u)-\bfbeta_0\bigr\}
-\sqrt{n}\bigl\{\hat\bfbeta(v)-\bfbeta_0\bigr\}\bigl|
> {\varepsilon \over 3|D(\bfbeta_0,t_0)|} \biggr) = 0.
}
This, together with the asymptotic expansion (\ref{eq:SP_asymp}), implies that the stochastic process \\
$\bigl\{ \sqrt{n} \{\hat S_1(u,t_0)-\hat S_0(u,t_0)\}:$
$~0< \tau_0 \leq u \leq \tau \bigr\}$ is asymptotically tight 
and converges weakly to the Gaussian process $\xi$.

%Bibliography
\bibliographystyle{apalike}  
\bibliography{main}

\end{document}